\documentclass[fleqn,usenatbib]{mnras}

\usepackage{mathptmx}

\usepackage[T1]{fontenc}

\DeclareRobustCommand{\VAN}[3]{#2}
\let\VANthebibliography\thebibliography
\def\thebibliography{\DeclareRobustCommand{\VAN}[3]{##3}\VANthebibliography}

\usepackage{graphicx}	
\usepackage{amsmath}	
\usepackage{amssymb}	
\usepackage{float}
\usepackage{booktabs}
\usepackage{hyperref}
\usepackage{subcaption}
\usepackage{mciteplus}
\usepackage{soul}
\usepackage{orcidlink}
\usepackage{xcolor}
\usepackage{ulem}
\usepackage{etoolbox}

\newcommand{\e}{\mathrm{e}}
\newcommand{\ii}{\mathrm{i}}

\newcommand{\rev}[1] {\ifmmode{\mbox{\color{black}$#1$}}\else{\color{black}#1}\fi}
\newcommand{\revdel}[1]{\ifmmode{\mbox{\color{black}\sout{$#1$}}}\else{\color{black}\sout{#1}}\fi}

\usepackage{xcolor}
\newcommand{\revb}[1]{\textcolor{black}{#1}}
\newcommand{\revdelb}[1]{\ifmmode{\mbox{\color{red}\sout{$#1$}}}\else{\color{black}\sout{#1}}\fi}
\usepackage{mhchem}

\usepackage{etoolbox}
\robustify{\ion}

\title[Depolarisation in the SMC]{Probing the Turbulent Magnetised Interstellar Medium of the Small Magellanic Cloud through Depolarisation with POSSUM Observations}

\author[J. M. Price et al.]{
Jason M. Price \orcidlink{0000-0001-7013-0562},$^{1}$\thanks{E-mail: jason.price1@anu.edu.au (JMP)} 
Yik Ki Ma \orcidlink{0000-0003-0742-2006},$^{1,2}$
Amit Seta \orcidlink{0000-0001-9708-0286},$^{1}$
N. M. McClure-Griffiths \orcidlink{0000-0003-2730-957X},$^{1}$
B. M. Gaensler \orcidlink{0000-0002-3382-9558},$^{3,4,5}$
\newauthor 
S. Lyla Jung \orcidlink{0000-0001-5512-3735},$^{6}$
Cameron L. Van Eck \orcidlink{0000-0002-7641-9946} ,$^{1}$
David McConnell \orcidlink{0000-0002-2819-9977},$^{1}$
Lindsey Oberhelman \orcidlink{0009-0009-2848-5269},$^{1}$
\newauthor
J. D. Livingston \orcidlink{0000-0002-4090-8000},$^{2}$
Takuya Akahori \orcidlink{0000-0001-9399-5331},$^{7}$ 
Shane P. O'Sullivan \orcidlink{0000-0002-3968-3051},$^{8}$
Denis Leahy \orcidlink{0000-0002-4814-958X},$^{9}$
C. S. Anderson \orcidlink{0000-0002-6243-7879}$^{1}$
\\
\\  
$^{1}$ Research School of Astronomy \& Astrophysics, The Australian National University, Canberra ACT 2611, Australia \\
$^{2}$ Max-Planck-Institut f\"ur Radioastronomie, Auf dem H\"ugel 69, D-53121 Bonn, Germany \\
$^{3}$ Department of Astronomy and Astrophysics, University of California, Santa Cruz, 1156 High Street, Santa Cruz, CA 95060, USA \\
$^{4}$ Dunlap Institute for Astronomy and Astrophysics, University of Toronto, 50 St.\ George Street, Toronto, ON M5S 3H4, Canada \\
$^{5}$ David A.\ Dunlap Department of Astronomy and Astrophysics, University of Toronto, 50 St.\ George Street, Toronto, ON M5S 3H4, Canada\\
$^{6}$  Department of Physics, University of Oxford, Keble Road, Oxford OX1 3RH, United Kingdom\\
${7}$  Mizusawa VLBI Observatory, National Astronomical Observatory Japan, 2-21-1 Osawa, Mitaka, Tokyo 181-8588, Japan \\
${8}$ Departamento de Física de la Tierra y Astrofísica \& IPARCOS-UCM, Universidad Complutense de Madrid, 28040 Madrid, Spain \\
$^{9}$ Department of Physics and Astronomy, University of Calgary, Calgary T2N 1N4, Canada \\
}

\date{Accepted 5/11/26. Received 7/04/26; in original form ZZZ}

\pubyear{2026}

\begin{document}
\label{firstpage}
\pagerange{\pageref{firstpage}--\pageref{lastpage}}
\maketitle

\begin{abstract}
The interstellar medium (ISM) of the Small Magellanic Cloud (SMC) provides a unique laboratory for studying magnetised gas in a low-metallicity, tidally disturbed environment. We investigate this magnetised turbulent medium using broad-band polarimetric data from the Australian Square Kilometre Array Pathfinder (ASKAP)’s Polarisation Sky Survey of the Universe’s Magnetism (POSSUM), leveraging a high-density Rotation Measure (RM) grid comprising \rev{951} RMs ($\sim24$ RMs per square degree) to probe the magneto-ionic structure of the SMC. We statistically compare the number densities of polarised and non-polarised non-thermal background sources on a per-square-degree basis, revealing spatially coherent depolarisation signatures that correlate with enhanced plane-of-sky RM fluctuations. This is consistent with a turbulent cascade from large-scale structures. Analysed through RM synthesis and Stokes $QU$-fitting, we attribute the depolarisation to a turbulent magneto-ionic medium within the telescope's beam ($\rev{\lesssim6}~\mathrm{pc}$). Faraday complexity is detected in $\sim$24\% of the RM \rev{sources}; however, no coherent spatial structure is observed in the complexity distribution, and we attribute this primarily to internal source structure or an extended halo surrounding the SMC rather than to its main body. Furthermore, \revb{as previously identified in the Large Magellanic Cloud, depolarisation in the SMC increases with emission measure; however, the underlying Faraday dispersion is $\sim$6 times weaker. With the thermal electron density smooth on the scales of the field fluctuations, this depolarisation is likely driven by magnetic field fluctuations.}
\end{abstract}




\section{Introduction}

The detailed structure of magnetic fields in the interstellar medium (ISM) of galaxies has largely remained elusive observationally due to the sparsity of high-sensitivity data. It is well established that magnetic fields thread through the ionised, atomic, and molecular phases of the ISM \citep{2015Natur.520..518Li, Ferriere2020, 2023.Ma, 2025.Seta} and influence the structure and flow of gas \citep{2012.Heiles, 2013.Gent, 2014.Clark,2016.Inoue, 2016.Evirgen, 2018.Villagran, 2018.Gazol, 2020.Bracco, SetaF2022, 2022.Campbell, 2023MNRAS.522.4161J}. Conversely, large-scale motions and turbulence within the ISM modify the properties of magnetic fields \citep{BrandenburgS2005, Rincon2019, SetaEA2020}. The complex dynamical interplay between the gas, turbulence, and magnetic fields generates ISM structures shaped by both small-scale processes, such as star formation and supernova explosions \citep{2006ApJ...652.1339M, 2006.Lyutikov, 2023.Hidalgo-Pinedazq, 2024.Jung, 2025A&A...694A.296G}, and large-scale processes, including galactic-scale shear and outflows, and dynamical interactions in merging systems \citep{ShukurovS2021, 2023.Brandenburg}.

Studies of magnetised structures in a variety of astrophysical scenarios have advanced rapidly with the increasing capabilities of modern radio telescopes. The Polarisation Sky Survey of the Universe’s Magnetism \citep[\href{https://possum-survey.org}{POSSUM};][]{Gaensler_2025} survey, underway with the Australian Square Kilometre Array Pathfinder \citep[ASKAP;][]{2021.Hotan}, is using Faraday rotation and polarised synchrotron emission to measure Galactic and extragalactic magnetic fields with an unprecedented combination of sensitivity, bandwidth, and sky area.

In this paper, we exploit POSSUM Pilot II ($800-1088~\mathrm{MHz}$) observations to study the magnetic structure of the ISM in the Small Magellanic Cloud (SMC). The SMC is a tidally disrupted dwarf satellite galaxy of the Milky Way (MW), located $62.4 \pm 0.8$ kpc away \citep{2020.Graczyk}. It is thought to be on its first passage around the MW, in a tidal interaction with its companion, the Large Magellanic Cloud (LMC). This interaction has given rise to several features, including the Magellanic Bridge \citep{2013.Nidever, 2018.Zivick}, the Leading Arm \citep{1998Natur.394..752P, 2008.Nidever}, and the Magellanic Stream \citep{1974.Mathewson, 2016.DOnghia}. The SMC itself comprises two primary regions: the Bar, which contains the majority of its gaseous and stellar mass, and the Wing, a tidal feature extending towards the Magellanic Bridge that connects the SMC and LMC. The SMC is characterised by low gas metallicity and low star formation rates \citep{2016.DOnghia}. Due to its proximity, the SMC is an excellent target for studying magnetic fields in an interacting dwarf galaxy \citep{1986.Haynes, 1970Mathewson.a, 2008.Mao, 2022MNRAS.510..260LivingstonSMC, SetaEA2023, 2023.Ma, 2024.Jung}. In this work, we primarily use the Faraday rotation effect to probe the magnetised ISM of the SMC.

Faraday rotation refers to the change in the observed angle of linear polarisation ($\psi$) from its intrinsic value at the source ($\psi_\mathrm{source}$) according to the relation: $\psi = \psi_\mathrm{source} + \mathrm{RM}\,\lambda^2$, where the degree of rotation is quantified by the Rotation Measure (RM) or Faraday depth ($\phi$)\footnote{For sources whose polarisation angle varies linearly with $\lambda^2$, the classical RM and Faraday depth are equivalent ($\phi = \mathrm{RM}$). Faraday depth is used to identify distinct magneto-ionic Faraday-rotating elements within the synthesised beam, and is typically used when synchrotron emission and Faraday rotation are spatially mixed. In this work, the complex magneto-ionic structure observed necessitates the use of Faraday depth, as a single RM cannot capture the full complexity. We refer to the brightest component in polarised intensity as $\phi_\mathrm{peak}$.} \citep{1966.Burn,2005.Brentjens, 2015.Klein} expressed as:
\begin{equation}\label{eqn: rm}
     \bigg(\frac{\phi}{\mathrm{rad} ~\mathrm{m}^{-2}} \bigg) = 0.812 \int_0^L \bigg(\frac{n_e}{\mathrm{cm}^{-3}}\bigg) \, \bigg(\frac{B_\parallel}{\mu\mathrm{G}}\bigg) \, \bigg(\frac{{\rev{\rm d} r}}{\mathrm{pc}}\bigg),
\end{equation}

\noindent where $n_e$ is the thermal electron density, $B_\parallel$ is the magnetic field strength along the line of sight (LOS), ${\rev{\rm d} r}$ is the differential path length element along the LOS, and $L$ represents the path length through the Faraday-rotating component; the sign of the $\phi$ represents the mean direction of the LOS magnetic field, positive towards the observer and negative away \citep[see][]{1979rpa..bookRybicki,2021.Ferrière}. The polarised intensity (PI) of the source, expressed as a complex vector in the Stokes $QU$ plane, is $\mathrm{PI}(\lambda^2) = Q(\lambda^2) + iU(\lambda^2)$, where the linear polarisation angle is given by \rev{$\psi = 0.5\tan^{-1}{(U/Q)}$}. When measured across a broad band, the polarised intensity enables characterisation of the magneto-ionic medium through which the emission has propagated.

However, variations in $B_\parallel$ and/or $n_e$ on spatial scales smaller than the beam can introduce heterogeneity in the Faraday rotation experienced by the polarised rays of light within the beam, leading to depolarisation. When these rays are convolved on the plane of sky (POS), the heterogeneity across their LOS Faraday rotation produces wavelength-dependent depolarisation (beam depolarisation). In the limit of infinitesimal angular resolution, all individual polarised rays would be fully resolved, and beam depolarisation would therefore vanish (for background sources). This depolarisation mechanism induce\rev{s} a random walk in the polarisation vector due to the stochastic Faraday rotation and is wavelength dependent \citep{1966.Burn,1998MNRAS.299..Sokoloff, 2015.Klein}. The standard deviation of Faraday depths realised during this random walk is quantified by the Faraday dispersion:
\begin{equation} \label{eqn: simga rm }
    \sigma_\phi = 0.812\,n_e\,B_\parallel \, \sqrt{l_0\,L}, 
\end{equation}
\noindent where $l_0$ is the \rev{correlation} length scale of the Faraday-rotating plasma \citep{2015.Klein, 2019.On, 2021.Seta_pul}. Under sufficiently strong Faraday dispersion, the polarised intensity of background sources may be suppressed below detectability \citep{1991.Tribble,1998MNRAS.299..Sokoloff, 2007.Stil,2014.Shneiderrj1, 2018.Machida}. While this leads to the loss of direct Faraday rotation information, the spatial distribution of these non-detections is itself a measure of Faraday dispersion regions, with fewer polarised sources \revb{indicating} stronger Faraday dispersion\revb{, assuming the sources are intrinsically polarised}, providing a powerful diagnostic of the turbulent magneto-ionic medium.

In this work, we present \rev{951} RMs ($\sim24$ RMs per square degree) across the field of the SMC, delivered by the pilot observations of the POSSUM survey, representing an approximately fivefold increase compared to previous studies \citep{2008.Mao, 2022MNRAS.510..260LivingstonSMC}. This dense ``RM grid” enables a detailed assessment of the SMC's magneto-ionic structure through an analysis of depolarisation. \rev{At the SMC's distance of 62.4~kpc \citep{2020.Graczyk}, our synthesised beam of $21''$ corresponds to a spatial resolution of $\approx6$~pc.}

This paper is organised as follows. Section~\ref{sec: 2 Observations} describes the POSSUM observations, data reduction procedures, and source-finding methodology. In Section~\ref{sec: 3 Methods}, we outline the methods used to assess depolarisation and Faraday complexity. Section~\ref{sec: 4 Results} presents the results, while Section~\ref{sec 5: Discussion} provides further analysis and discusses these findings in the context of the prior observations. We conclude in Section~\ref{sec: 7 summary}.

\section{Observational data}\label{sec: 2 Observations}
\subsection{POSSUM Observations of the SMC}

Spectro-polarimetric observations were made on the 7th August 2022 with a 10-hour integration time. The observations followed POSSUM's Band 1 specifications covering 288 MHz spanning $800-1088~\mathrm{MHz}$ with a $1~\mathrm{MHz}$ sampling, over approximately 40 square degrees of sky. We achieved a full-band sensitivity of $27~\mu\mathrm{Jy}~\mathrm{beam}^{-1}$ in Stokes $I$ and $17~\mu\mathrm{Jy}~\mathrm{beam}^{-1}$ in Stokes $Q$ and $U$, respectively. Upon inspection of our Stokes $I$ multi-frequency synthesis (MFS) image, substantial diffuse continuum emission from within the SMC was apparent. To prevent the obscuration of background sources by the SMC during our source-finding procedures, we subtracted its diffuse Stokes $I$ emission from the Stokes $I$ image using a median filter method \cite[described in Section~3.1.1 of][]{2024.Vanderwoude}, achieving a sensitivity of $27~\mu\mathrm{Jy}~\mathrm{beam}^{-1}$ in the MFS map. Subsequently, we performed source-finding in the resulting Stokes $I$ image using the pyBDSF (version 1.13.0) package \citep{2015ascl.soft02007M}, and identified 31,164 sources based on pixel brightness.

\begin{figure}
    \centering
    \includegraphics[width = \columnwidth]{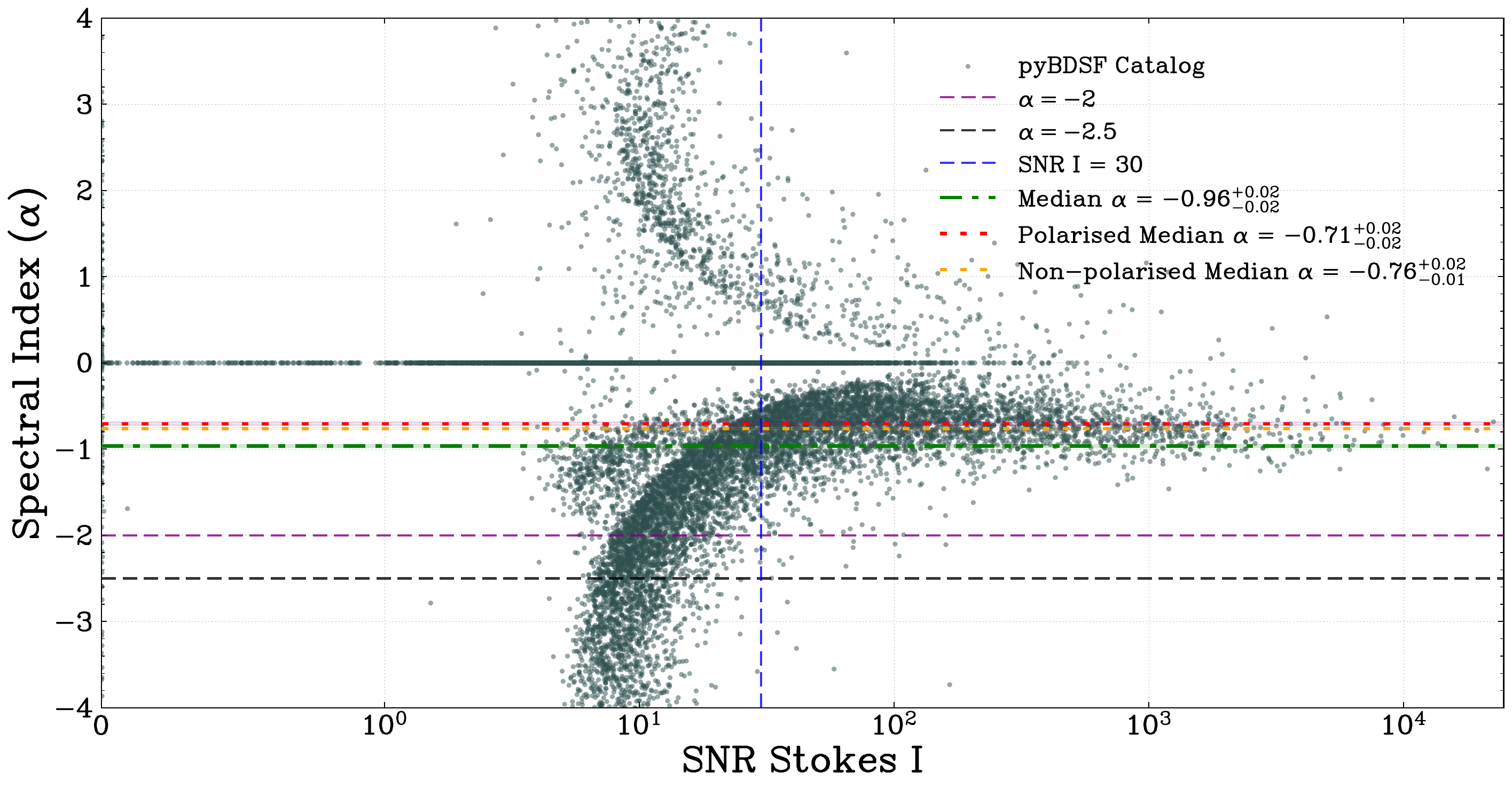}
    \caption{Spectral index ($\alpha$) versus \revb{Stokes $I$ signal-to-noise ratio (SNR)} fit by the POSSUM pipeline (Van Eck et al., in preparation). The majority of sources beyond the Stokes $I$ \revb{SNR} of 30 (blue, dashed vertical line) fall within the expected spectral index range of non-thermal sources. Median spectral index of the full, polarised and non-polarised distributions are shown as green, red and yellow lines respectively \rev{ (see Section~\ref{sec: Source selection} for further details)}. The uncertainties were determined via bootstrap resampling (1000 iterations), shown as the shaded region around the lines. Sources showing identically zero spectral index (totalling 16,783) are excluded from our analysis.}
    \label{fig: Spectral index dsitbruion of the whole catalouge}
\end{figure}

The Stokes cubes are convolved to a common beam of $21''$ ($\approx 6$~pc) over all frequency channels and then processed through the POSSUM pipeline (Van Eck et al., in preparation). The pipeline extracted Stokes $I$, $Q$, and $U$ values across the entire frequency range for each of the 31,164 sources identified above, with the diffuse polarised emission subtracted from each source using an annulus median filter \citep{oberhelman24}. This step guarantees that our measured emission from the background sources is not contaminated by foreground diffuse emission, which can induce Faraday complexity\footnote{A source is described as ``Faraday complex'' when its polarisation behaviour as a function of $\lambda^2$ is inconsistent with a single source of polarised emission Faraday-rotated by a uniform foreground $B_\parallel$ and $n_e$ element \citep{10.1093/mnras/268.2.497, 2015.Sun, 2012.O’Sullivan, 2021.Alger}.} and/or depolarisation\footnote{We confirmed the absence of foreground diffuse polarised emission \rev{from the SMC} by performing 3D RM synthesis on the Stokes cubes, which produced a Faraday depth spectrum for each pixel. The resulting peak Faraday depth map showed no large-scale coherent structure, \rev{indicating no significant SMC-associated diffuse polarised emission across the field}.} 
\citep{2024A&A...686A.104Ranchod}. The \rev{median} full-band RMS after diffuse emission subtraction, across all 31,164 sources, is \rev{27.5, 19.8, and 19.6} $\mu$Jy/beam in Stokes $I$, $Q$ and $U$, respectively. The pipeline then applies RM synthesis \citep{2005.Brentjens, Heald.2009} to every source using the {\sc \href{https://github.com/CIRADA-Tools/RM-Tools}{\texttt{RM-TOOLS}}} (version 1.3.0) software \citep{vaneck2026rmtoolssoftwareanalyzingpolarized}. For the aforementioned band and channel width, we achieved a maximum detectable scale in Faraday space of $\approx 41~\mathrm{rad}~\mathrm{m}^{-2}$, a maximum Faraday depth $\approx 8106~\mathrm{rad}~\mathrm{m}^{-2}$, and resolution (FWHM) of $\approx 59~\mathrm{rad}~\mathrm{m}^{-2}$ \citep{2005.Brentjens, 2019.Dickey}. \rev{We sampled the Faraday dispersion function over $\phi \in [-6000, 6000]~\mathrm{rad~m^{-2}}$ with ten samples across the \revb{Rotation Measure Transfer Function (RMTF)} FWHM, using inverse-variance channel weighting. The Stokes $I$ spectrum was modelled as a power law following the convention $I_\nu \propto \nu^{\alpha}$, with the polynomial order (zero, one, or two) selected per source by the Akaike Information Criterion (AIC). The resulting spectral indices for the full catalogue are shown in Figure \ref{fig: Spectral index dsitbruion of the whole catalouge}.} \revb{Within the catalogue, 16,783 sources had a reported spectral index of identically zero, which indicates that the POSSUM pipeline's first-order fit to the Stokes $I$ spectrum did not significantly improve the AIC compared to the zeroth-order fit, and were excluded from our analysis.}

\subsection{Neutral and Ionised Gas Density Tracers}\label{sec: Methods 3: Neutral and Ionised Gas Density Tracers}

To obtain magnetic field information from $\phi$, details about thermal electron density are required. Usually, this is provided via H$\alpha$ and/or \rev{neutral hydrogen (\ion{H}{I})} data. Here, we use \ion{H}{I} data to define the \rev{extent} of the SMC and H$\alpha$ to extract thermal electron density information.

Neutral hydrogen column density measurements, $N_\mathrm{\ion{H}{I}}~(\mathrm{cm}^{-2})$, are taken from the ASKAP \ion{H}{I} observations presented by \citet{2022.Pingel} with a spectral resolution of $\rev{\sim}~0.98~\mathrm{km~s^{-1}}$ and angular resolution of $\rev{\sim}~30''$. The spatial coverage of these data is shown in Figure~\ref{fig: RMs from QU fitting in the sky}. Each background source is assigned an $N_\mathrm{\ion{H}{I}}$ value using a nearest-pixel sampling approach in the \ion{H}{I} data cube. For sources lying outside the ASKAP \ion{H}{I} footprint or with spurious values at the beam edge, $N_\mathrm{\ion{H}{I}}$ values are instead taken from the GASS \ion{H}{I} survey \citep{2009.McClure-Griffiths}.

We derive the Emission Measure (EM) from H$\alpha$ observations. We use two sets of H$\alpha$ observations, each at \rev{a different resolution}. The first dataset is sourced from the Wisconsin H-Alpha Mapper (WHAM) survey, \rev{which observes diffuse H$\alpha$ emission in} the SMC at an angular resolution of $3600''$ \citep{2019.Smart} and covers the entire POSSUM field. The second is from the Magellanic Cloud Emission-Line Survey (MCELS), at a higher angular resolution of $5''$ and covering a central region of approximately $4.5^\circ \times 3.5^\circ$, capturing \rev{more} small-scale ionised structures \rev{compared to the WHAM survey} \citep{2015.Winkler.MCELS, 1999.Smith.MCELS}. Both images are corrected for internal extinction by the SMC and the MW following \cite{2013.Barger} and \cite{2019.Smart}. For polarised sources within the MCELS field of view, we assign associated emission from the MCELS data unless the data are determined to be non-physical (negative). In such cases, WHAM measurements are utilised instead. This captures a large range of extinction-corrected H$\alpha$ intensity ($I_{\mathrm{H}\alpha,\mathrm{corr}}$) observed across the SMC. The Emission Measure (EM) is then calculated as:
\begin{align}
    \Bigg(\frac{\mathrm{EM}}{\mathrm{cm}^{-6} \mathrm{pc}}\Bigg) &= \Bigg(\frac{\langle n_e \rangle}{\mathrm{cm}^{-3}\mathrm{pc}}\Bigg)^2~ \Bigg(\frac{L_{H^+}}{\mathrm{pc}}\Bigg) \nonumber \\
    &\approx 2.75~\Bigg(\frac{I_{\mathrm{H}\alpha,\mathrm{corr}}}{\mathrm{R}} \Bigg)~\Bigg(\frac{T_e}{10^4 \mathrm{K}}\Bigg)^{0.942},
\end{align}
where $T_e$ is the electron temperature, assumed to be $10^4~\mathrm{K}$ \citep{2019.Smart} and $L_{H^+}$ is the path length of the ionised medium \rev{\citep[see Equation~4 of][for further details]{2013.Barger}}.

\section{Methods}\label{sec: 3 Methods}

To characterise the $\phi$ and Faraday complexity of each polarised source, we employed two discrete techniques applied to the observed Stokes $Q$ and $U$ polarisation spectra. First, we performed RM synthesis as implemented \rev{in} the POSSUM pipeline, together with appropriate metrics to examine Faraday complexity from the \rev{RMCLEAN-reconstructed} Faraday spectrum \citep{Heald.2009}. Second, we conducted a parametric fit of theoretical models to the observed polarimetric spectra using $QU$-fitting \citep{2012.O’Sullivan, 2015.Sun, 2016.Anderson, 2018.Pasetto, 2019.Miyashita, 2021.Thomson}. Further analysis of depolarisation was explored in a statistical spatial framework.

\subsection{RM Synthesis: Faraday Complexity Metrics}\label{sec: rm synth Faraday Complexity Metrics}

Currently, two primary metrics exist to classify a source as Faraday complex from its Faraday spectrum, $F(\phi)$. First, the second moment (M$_2$) of the RMCLEANed Faraday spectra provides an indication of how wide the spectrum is \citep{Brown_POSSUM11, 2015.Anderson, 2017.Purcell.West, 2019.Dickey, 2021MNRAS.Livingston.Heightened.Faraday.complexity, 2023.Thomson}. We normalise M$_2$ by the FWHM of the RMTF to account for the resolving power in Faraday depth space and use this new value ($\mathfrak{m}_{2,\mathrm{norm}}$) throughout the paper \citep{2011.Farnsworth, 2014.Kumazaki, 2015.Sun, 2023.Thomson}. \rev{We adopt a threshold of $\mathfrak{m}_{2,\mathrm{norm}} > 0.5$ to classify a source as Faraday complex \citep{2024.Vanderwoude}.}

The second metric of Faraday complexity is $\sigma_{\mathrm{add}}$, developed by \cite{2017.Purcell.West} in pursuit of a more robust and discriminative measure of complexity. It is determined by passing the observed Stokes $Q$ and $U$ spectra through RM synthesis to generate a Faraday spectrum, extracting $\phi, \psi$ \rev{and the fractional polarisation $p = \mathrm{PI}/I$}. These quantities are passed into a Faraday simple (thin) model generating synthetic Stokes $Q$ and $U$. The synthetic spectrum is subtracted from the observed and structure beyond Gaussian noise in the residuals between them indicates a source is Faraday complex. A threshold of $\sigma_{\mathrm{add}} \geq 1$ is used here to identify Faraday complex structure in the residuals. This complexity metric is a POSSUM pipeline product, and is detailed by \cite{2017.Purcell.West}, \cite{2024.Vanderwoude} and \cite{vaneck2026rmtoolssoftwareanalyzingpolarized}.

\subsection{Stokes $QU$-Fitting} 
\label{sec: QU fitting methods}

Stokes $QU$-fitting is a model-fitting technique that estimates the magneto-ionic structure along the line of sight by comparing observed Stokes $Q(\lambda^2)$ and $U(\lambda^2)$ with modelled counterparts. To understand this approach, consider how the plane of polarisation interacts with various astrophysical environments. At one wavelength, the ellipticity of the light is characterised by the Stokes parameters $I$, $Q$, $U$, and $V$. The Stokes $Q$ and $U$ parameters define a polarisation vector in the $QU$ plane, the magnitude of which gives the linear polarised intensity ($\mathrm{PI}$). Stokes $QU$-fitting models how the plane of polarisation changes across the band. For Faraday complex sources experiencing depolarisation, the polarisation vector's magnitude diminishes towards lower \rev{frequencies}.

\subsubsection{$QU$-Fitting Models}

Models chosen for fitting were selected based on the most probable types of astrophysical environments resolvable with these data. We expect some manifestation of Faraday complexity either local or intrinsic to the source \citep{2015.Anderson, 2017.O'Sullivan}. The models are labelled by their \texttt{RM-TOOLS} designation for consistency, and are described below. An isotropic homogeneous magnetised plasma in the foreground inducing Faraday rotation is classified as Faraday thin and is represented by model 1 (m1). When turbulence is present in the foreground magnetised plasma, the source is described by model 2 (m2). Model 3 (m3) describes two distinct polarised emission components subjected to the same level of turbulence in the foreground plasma. Model 4 (m4) describes a scenario where two emission components are subjected to different levels of turbulence in the foreground. Model 11 (m11) describes two separate regions of homogeneous magnetised plasma inducing two distinct modes of Faraday rotation, whilst Model 111 (m111) depicts three regions. The functional form of these models, their Faraday complexity classification and the number of free parameters ($n_{\mathrm{f}}$), are given in Table~\ref{tab: QU-Fitting models fucntions }. 

\begin{table*}
\resizebox{\textwidth}{!}{%
\begin{tabular}{@{}lllcc@{}}
\toprule
Model & Description & Function ($P_{\mathrm{mod}}(\lambda^2; \theta)$) & $n_{\mathrm{f}}$  &  \\ \midrule
m1  & Faraday thin &
$p_{0}\,\e^{2\ii(\psi_{0} + \phi \lambda^2)}$ & 3 & \\

m2  & External Faraday Dispersion &
$p_{0}\,\e^{-2\sigma_{\phi}^2 \lambda^4}\,\e^{2\ii(\psi_{0} + \phi \lambda^2)}$ & 4 & \\

m3  & Two components with common external Faraday dispersion &
$\big[p_{0,1}\,\e^{2\ii(\psi_{0,1} + \phi_{1}\lambda^2)} + p_{0,2}\,\e^{2\ii(\psi_{0,2} + \phi_{2}\lambda^2)}\big]\,\e^{-2\sigma_{\phi}^2 \lambda^4}$ & 7 & \\

m4  & Two components with separate external Faraday dispersion &
$p_{0,1}\,\e^{2\ii(\psi_{0,1} + \phi_{1}\lambda^2)}\,\e^{-2\sigma_{\phi,1}^2 \lambda^4}
+ p_{0,2}\,\e^{2\ii(\psi_{0,2} + \phi_{2}\lambda^2)}\,\e^{-2\sigma_{\phi,2}^2 \lambda^4}$ & 8 & \\

m11 & Two Faraday thin &
$p_{0,1}\,\e^{2\ii(\psi_{0,1} + \phi_{1}\lambda^2)} + p_{0,2}\,\e^{2\ii(\psi_{0,2} + \phi_{2}\lambda^2)}$ & 6 & \\

m111 & Three Faraday thin &
$p_{0,1}\,\e^{2\ii(\psi_{0,1} + \phi_{1}\lambda^2)} + p_{0,2}\,\e^{2\ii(\psi_{0,2} + \phi_{2}\lambda^2)} + p_{0,3}\,\e^{2\ii(\psi_{0,3} + \phi_{3}\lambda^2)}$ & 9 &
\\ \bottomrule
\end{tabular}%
}
\caption{$QU$-fitting models; model number corresponding to {\sc \href{https://github.com/CIRADA-Tools/RM-Tools}{\texttt{RM-TOOLS}}} \citep{vaneck2026rmtoolssoftwareanalyzingpolarized} designation, model description, functional form and number of free parameters ($n_{\mathrm{f}}$), where $p_0$ is the intrinsic degree of fractional polarisation, $\psi_0$ is the intrinsic polarisation angle, and $\phi$ is the Faraday depth component.}
\label{tab: QU-Fitting models fucntions }
\end{table*}

\subsubsection{Best-Fit Model Selection}\label{sec: best fit model selection}

The parameter space of these models was searched with the nested sampling technique \citep{2004.Skilling, 2020.Albert}, with parameters determined through the minimisation of $\chi^2$, considering the observed polarised spectrum $P_{\mathrm{obs}}(\lambda^2)$ and the model polarised spectrum $P_{\mathrm{mod}}(\lambda^2; \theta)$ for $\theta$ models parameters \citep{2019.Miyashita}. Uniform priors were adopted for all parameters: $0.001 \leq p_{0,i} \leq 1$, $0^{\circ} \leq \psi_{0,i} \leq 180^{\circ}$, $-500 \leq \phi_i \leq 500~\mathrm{rad\,m^{-2}}$, and $0 \leq \sigma_{\phi,i} \leq 30~\mathrm{rad\,m^{-2}}$. For multi-component models, additional constraints were imposed: $\sum_i p_{0,i} \leq 1$, and $0 \leq \Delta\phi_{i,j} \leq 1000~\mathrm{rad\,m^{-2}}$ between each pair of Faraday depth components.

To exclude any non-physical or ``$QU$-fitting biased'' results, the model fits of m1, m2 and m11 were scrutinised and filtered in accordance with the accurate model parameter spaces identified for the POSSUM band 1 data by \cite{2026MNRAS.547ag394O}, while models m3, m4, m111 were filtered in accordance with \cite{Ma.2019I}. These filters ensure all valid solutions satisfy the following criteria: any component has an \revb{PI SNR} $> 7$, all Faraday depth components in multiple component models are separated by $> 10~\mathrm{rad~m}^{-2}$, and any component in Faraday dispersion models has SNR in $\sigma_\phi > 3$. From these valid solutions, we selected the best model based on Bayesian Evidence ($EVD$), choosing the model with the largest $EVD$ value \citep{2026MNRAS.547ag394O}.

\subsection{Statistical Depolarisation}\label{sec: Methods; Statistical Depolarisation}

We examined the spatial distribution of polarised sources statistically by subdividing the field into $1.5\times1.5=2.25$ square degree bins, corresponding to $1635~\mathrm{pc} \times 1635~\mathrm{pc}$, with CASA\footnote{The Common Astronomy Software Applications (\href{https://casa.nrao.edu}{CASA}-- version 6.5.4-9) software is a tool kit developed by NRAO for radio observation and is an open source application. We use the function (task) \href{https://casadocs.readthedocs.io/en/stable/api/tt/casatasks.analysis.imrebin.html}{imrebin} for our sub-division.}. Within these bins, we calculate the fraction of polarised \revb{sources} per bin as:
\begin{equation} \label{eqn: DP frac ratio}
    \revb{f_\mathrm{pol}} = \frac{N_\mathrm{pol}}{N_{\mathrm{tot}}},
\end{equation}
where $N_\mathrm{pol}$ represents the number of polarised sources per bin and $N_{\mathrm{tot}}$ the total sources (both polarised and non-polarised) per bin. Assuming the source density adheres to a Poisson distribution, the error associated with each bin is calculated as $\sqrt{N}$, where $N$ denotes the number of sources, whether polarised or total. This error is then propagated to estimate the uncertainty in $\revb{f_\mathrm{pol}}$ as:
\begin{equation}
    \revb{\sigma_{f_{\mathrm{pol}}}} =  \revb{f_{\mathrm{pol}}}\sqrt{\left(\frac{\sigma_{\mathrm{pol}}}{N_{\mathrm{pol}}}\right)^2 + \left(\frac{\sigma_{\mathrm{tot}}}{N_{\mathrm{tot}}}\right)^2}.
\end{equation}
This error assumes no correlation between the source densities and is thus the upper limit of the error. Only bins containing more than five polarised sources are considered to yield a reliable estimate of the uncertainty. We therefore exclude bins with $\leq 5$ polarised sources from our analysis (Figure~\ref{fig: D vs sigma RM} and \ref{fig: D and STD-RM with column densities}). Additionally, we calculate the dispersion of peak Faraday depth (RM) per bin as,
\begin{equation}
    \Sigma_{\phi_\mathrm{peak}} = \sqrt{\frac{\sum{(\phi_\mathrm{peak} - \overline{\phi_\mathrm{peak}})^2}}{N_\mathrm{pol}-1}} ~(\mathrm{rad~m}^{-2}).
\end{equation}
This is the standard deviation of peak Faraday depth per bin, where $ \overline{\phi_\mathrm{peak}}$ is the mean peak Faraday depth per bin. We repeat these calculations with bin sizes of $0.5\times0.5=0.25$ and $1.0\times1.0=1.0$ square degrees and find similar results; these are discussed in Appendix~\ref{sec: Appendix sub figs}. \revb{Subscripts on $f_{\mathrm{pol}}$, $\sigma_{f_{\mathrm{pol}}}$ and $\Sigma_{\phi_\mathrm{peak}}$ denote either the bin area in deg$^2$ (e.g.\ $f_{\mathrm{pol},2.25}$) or a specific bin as labelled in Figure~\ref{fig: Depaolrisation map per square degree} (e.g.\ $\Sigma_{\phi_\mathrm{peak},\mathrm{D2}}$).} \rev{We do not correct the observed Faraday depths for the Galactic foreground; we instead assume the Galactic RM is approximately constant on the scales of our analysis bins. From the STAPS Milky Way RM reconstruction \citep{2025A&A...694A.169Sun, 2025A&A.Raycheva}, the residual foreground variation across our bins is $\sim 2.6~\mathrm{rad~m}^{-2}$, which is well below the statistical uncertainty on our on-SMC RM dispersion ($\approx 9~\mathrm{rad~m}^{-2}$).}
To demonstrate the statistical significance of the observed spatial structure, we performed a Monte Carlo spatial randomisation test. \rev{We uniformly redistributed all source positions across the field 10,000 times, preserving each source's Faraday depth but randomising their locations; the total source count remains unchanged.} For each iteration, we recalculated $\revb{f_\mathrm{pol}}$ and $\Sigma_{\phi_\mathrm{peak}}$ in each bin.

\section{Results}\label{sec: 4 Results}

\subsection{Polarised and Non-Polarised Source Selection}\label{sec: Source selection}

To ensure that we worked with a reliable catalogue of radio sources, we excluded sources with a Stokes $I$ SNR below 30 (see Figure~\ref{fig: Spectral index dsitbruion of the whole catalouge}), spectral index errors exceeding $0.5$, spectral indices below $-3$, or fewer than 200 frequency channels.

\rev{From this parent sample, we constructed two sub-catalogues, polarised and non-polarised.} We classified sources as polarised if they had a \revb{PI SNR} greater than 7 and a fractional polarisation ($p$) \rev{between 0.01 and 0.72, with the lower bound set by the residual instrumental polarisation and the upper bound by the maximum theoretical fractional polarisation achievable for synchrotron emission \citep{1979rpa..bookRybicki}}. These criteria yielded a total of \rev{$1076$} polarised sources \rev{from} the RM synthesis results. However, our $QU$-fitting analysis (see \rev{S}ection \ref{sec: QU-Fitting Results}) \rev{revealed} that \rev{116} of these sources did not possess a model fit satisfying the criteria for a successful $QU$-fitting fit. So, we also excluded these from further analysis. An additional nine anomalous sources \rev{excluded from analysis} possessed spectral structure inadequately modelled by $QU$-fitting (see Section~\ref{sec: Unique sources} for a further discussion on these). 

\rev{To ensure that the non-polarised catalogue traced the background extragalactic synchrotron population and not non-polarised sources intrinsic to the SMC,} we removed 18 compact \ion{H}{II} regions identified by \cite{2012SerAJ.185...53Wong} and three planetary nebulae identified by \cite{2024.Cotton} within $21''$. To further isolate background extragalactic sources, we \rev{applied additional cuts removing sources with inverted spectra ($\alpha > 0.5$), typically associated with optically thick thermal emission from compact \ion{H}{II} regions \citep{2007prpl.Hoare, 2016era.Condon}. We additionally removed sources with substantially extended morphologies (peak-to-total flux ratio $< 0.7$), which captures optically thin thermal sources ($\alpha \approx -0.1$) that are characteristically extended \citep{2019A&A...627A.175Medina}.}

\rev{We then removed duplicate detections both within and between the two catalogues using a $21''$ matching threshold, retaining the source with the highest signal-to-noise ratio in the polarised intensity for polarised sources, and in Stokes $I$ for non-polarised sources. Where a source appeared in both catalogues, it was kept only in the polarised catalogue.}

\begin{figure*}
    \centering
    \includegraphics[width = \textwidth]{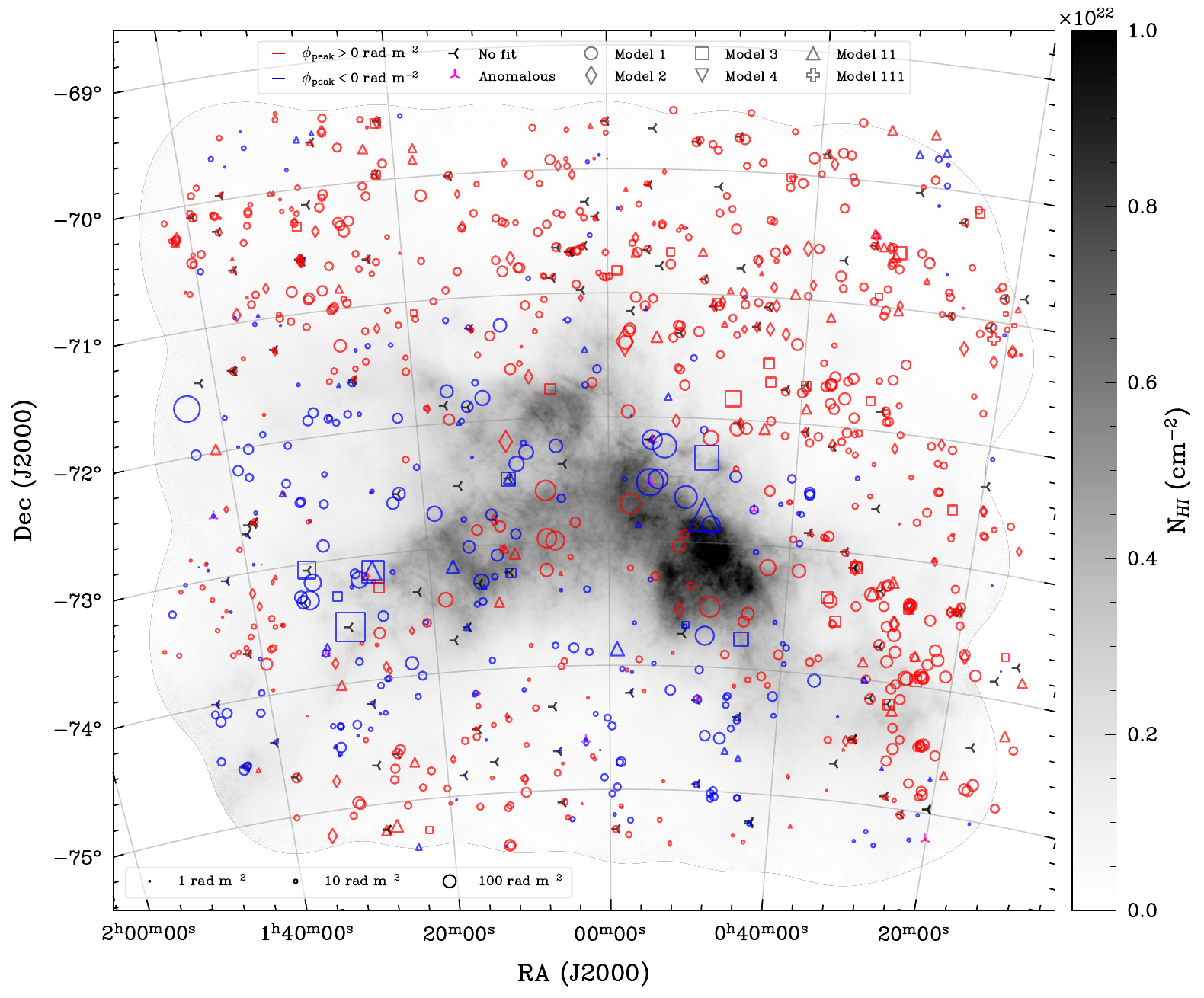}
    \caption{RM grid across the SMC from the POSSUM Pilot II observations processed through the POSSUM pipeline, totalling \rev{951} RMs. The area of the markers are linearly scaled by the absolute magnitude of the peak Faraday depth ($\phi_\mathrm{peak}$), the largest at $419~\mathrm{rad~m}^{-2}$, coloured by the mean direction of the LOS component of the magnetic field, with red indicating towards us (positive RM) and blue away from us (negative RM). Markers of the RM sources are separated by Stokes $QU$-fitting model type, whilst left leaning tri markers represent polarised sources with no model fit, discussed in Section~\ref{sec 5: Discussion}. Upwards leaning tri markers represent nine sources designated as Faraday simple by $QU$-fitting but which show anomalous spectral structure (see Section~\ref{sec: Unique sources}). The RMs in this image and work are not corrected for the Milky Way RM foreground. The background greyscale map shows the \ion{H}{I} column density of the SMC from the Pilot Survey of the Galactic ASKAP \ion{H}{I} (GASKAP-\ion{H}{I}) survey \citep{2022.Pingel}.}
    \label{fig: RMs from QU fitting in the sky}
\end{figure*}

\begin{figure}
    \centering
    \includegraphics[width=\columnwidth]{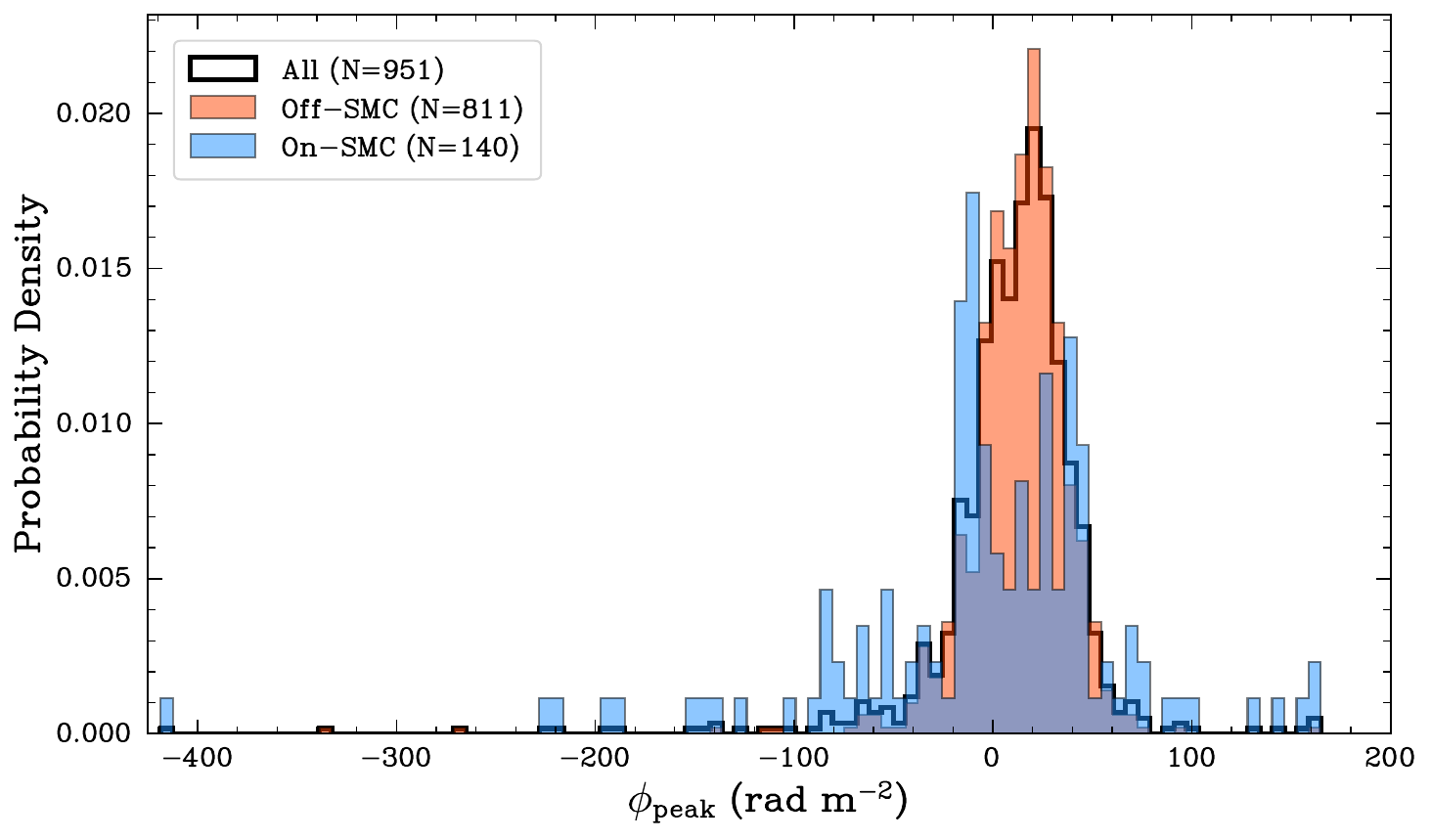}
    \caption{Histogram of peak Faraday depths separated by \ion{H}{I} column density. The full polarised catalogue is shown as the black line. Sources considered off-SMC—those probing \ion{H}{I} column densities less than $2\times 10^{21} ~\mathrm{cm}^{-2}$ are shown in orange, while sources considered on-SMC, probing higher \ion{H}{I} column densities, are shown in blue. These source are not Milky Way foreground corrected. A Kolmogorov–Smirnov test comparing the on-SMC and off-SMC distributions yields a statistic of 0.30 with a p-value of $\rev{5.32}\times 10^{-10}$. Distribution statistics are shown in Table \ref{tab: Model results and M2 median and mean}.}
    \label{fig: histogram of RM on/off SMC}
\end{figure}

This yielded a final polarised catalogue of \rev{951} sources, displayed in Figure~\ref{fig: RMs from QU fitting in the sky}, separated by their $QU$-fitting model assignment. This equated to approximately 24 RMs per square degree, a factor of $\sim\revb{22}$ improvement from the NVSS \citep{2009ApJ...702.1230T} and a factor of 5 from SPICE-RACS \citep{2026arXiv260516917T}. \rev{This RM density is comparable to that achieved in other POSSUM works \citep{2021.Anderson, 2024.Vanderwoude}.} Figure~\ref{fig: histogram of RM on/off SMC} shows a histogram of the peak Faraday depth ($\phi_\mathrm{peak}$) for these polarised sources, separated into off- and on-SMC regions using an \ion{H}{I} column density threshold of $2\times 10^{21} ~\mathrm{cm}^{-2}$.

\begin{figure}
    \centering
    \includegraphics[width=\columnwidth]{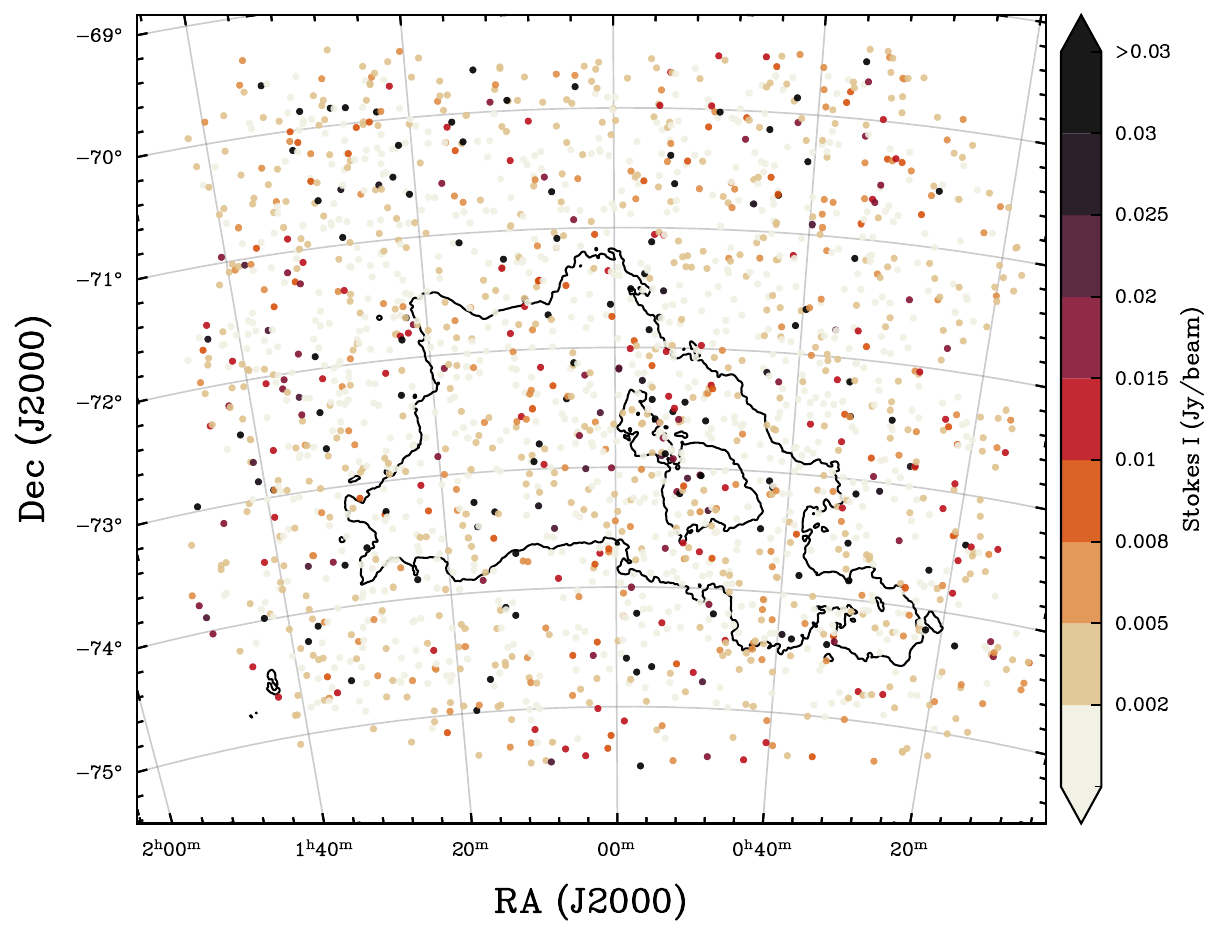}
    \caption{Locations of \rev{1919} non-polarised background sources, coloured by peak brightness. Contours show \ion{H}{I} column density ($\mathrm{N}_{\mathrm{\ion{H}{I}}}$) at 2 and 7 $\times 10^{21}$ cm$^{-2}$ from GASKAP-\ion{H}{I} \citep{2022.Pingel}. }
    \label{fig: unpoalrised sorces on the sky}
\end{figure}

The resulting \rev{1919} non-polarised sources are shown in Figure~\ref{fig: unpoalrised sorces on the sky}, exhibiting a median spectral index of $-0.76\pm{0.01}$. Previous observations of this nature suggest that the majority of these sources are unresolved Active Galactic Nuclei (AGN) or a single-lobe component \citep{2021.Norris, 2021.Filipovi,2024.Cotton, 2025SerAJ.211....9Khattab}. We characterised source size using the Stokes $I$ flux density distribution. Sources were classified as compact when $I_\mathrm{peak} / I_\mathrm{int} \ge 0.9$, and as extended otherwise. Table~\ref{tab: Model results and M2 median and mean} \rev{lists the compact fraction for each $QU$-fitting sub-population; overall, 75\% of the polarised catalogue is compact with this definition.}

\rev{We assessed the potential for per-channel (intra-channel) depolarisation arising from polarisation angle rotation across the finite width of a single frequency channel ($\Delta\nu = 1~\mathrm{MHz}$). This effect is largest for the highest-RM sources at the lowest observed frequency, where the change in $\lambda^2$ across a channel is greatest. Evaluating this worst case at the band edge ($800~\mathrm{MHz}$), the rotation across a single channel is $\Delta\psi = \lvert\mathrm{RM}\rvert\,\Delta(\lambda^2)_\mathrm{channel}$, and the largest value across the polarised catalogue ($\lvert\mathrm{RM}\rvert= 418.7~\mathrm{rad~m}^{-2}$) is $\Delta\psi = 0.15~\mathrm{rad}$ ($8.6^\circ$). Per-channel depolarisation is therefore negligible for all sources \citep{1966ARA&A...4..Gardner}.}

\subsection{$QU$-Fitting Results}\label{sec: QU-Fitting Results}

\begin{figure}
    \centering
    \includegraphics[width=\linewidth]{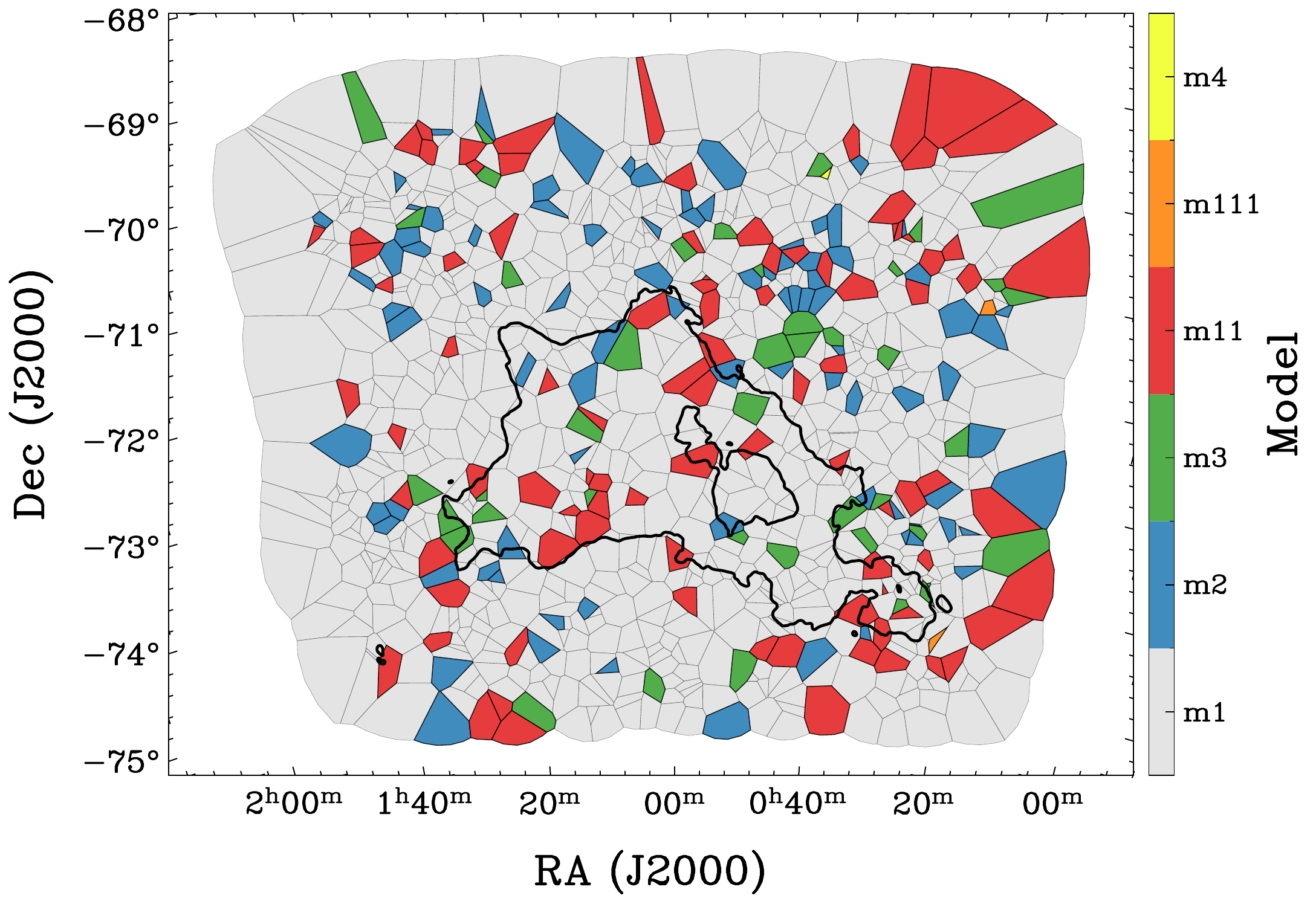}
    \caption{Spherical Voronoi tessellation of polarised sources across the SMC, colour separated by their best fit $QU$-fitting model. Tessellation was constructed with \texttt{scipy.spatial.SphericalVoronoi}, with cell boundaries traced as great-circle arcs on a unit celestial sphere. To ensure finite and well-behaved cells near the edge of the field, artificial boundary points are placed just outside the analysis region and excluded from visualisation. Contours show \ion{H}{I} column density ($\mathrm{N}_{\mathrm{\ion{H}{I}}}$) at 2 and 7 $\times 10^{21}$ cm$^{-2}$ from GASKAP-\ion{H}{I} \citep{2022.Pingel}. The tessellation highlights the spatial structure of Faraday complexity in the field. Each cell represents only the distance to the nearest source, not a region of influence; a length scale for $\phi$ should therefore not be inferred from this figure.}
    \label{fig: Voronoi Faraday complexity}
\end{figure}

We applied the six different $QU$-fitting models in Table~\ref{tab: QU-Fitting models fucntions } to the \rev{1076} polarised sources identified through RM synthesis. All sources were successfully fitted with a model. However, \rev{116} sources failed to meet our reliability thresholds and were excluded from the catalogue and subsequent analysis (see Section~\ref{sec: best fit model selection}). We classified the remaining sources as either Faraday simple (\rev{727} sources), Faraday complex (\rev{224} sources) or anomalous (9 sources); we discuss anomalous sources in Section~\ref{sec: Unique sources}. Approximately 24\% of the polarised catalogue exhibits Faraday complexity, with 14\% of the polarised catalogue exhibiting Faraday dispersion ($\sigma_\phi$). The distribution of the model fits and statistics on the populations is detailed in Table~\ref{tab: Model results and M2 median and mean} and shown across the SMC as a spherical Voronoi tessellation in Figure~\ref{fig: Voronoi Faraday complexity} and \rev{an} RM grid-type structure in Figure~\ref{fig: RMs from QU fitting in the sky}. We found no statistical correlation between Faraday complexity and \ion{H}{I} column density or EM, which we discuss further in Appendix~\ref{sec: Farady complexity discussion}.

\subsection{Spatial RM Dispersion and Statistical Depolarisation}\label{sec: Spatial RM Dispersion and Statistical depol results}

\begin{figure*}
    \centering
    \includegraphics[width=16cm, height=10.9cm, keepaspectratio]{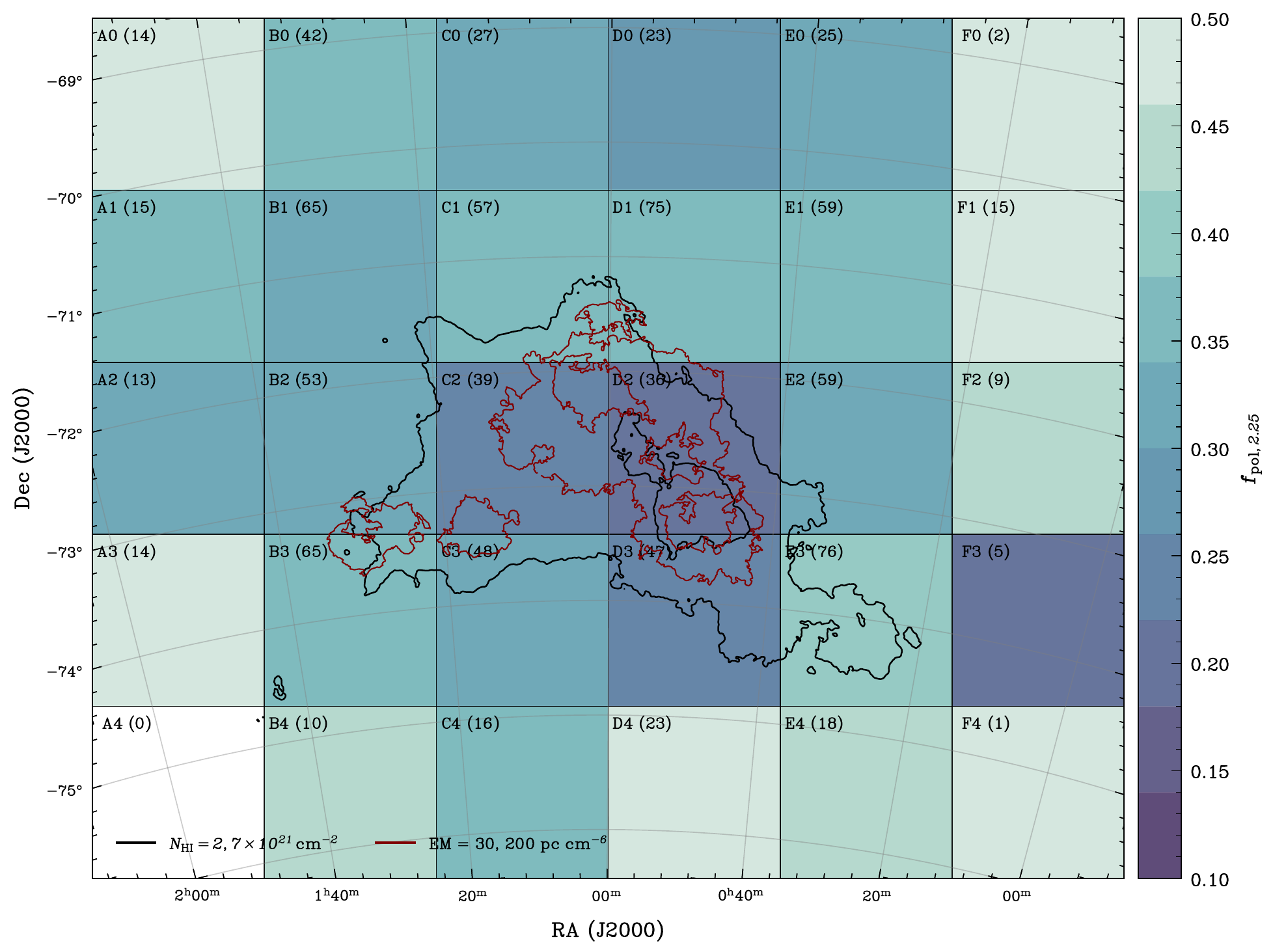}
    \includegraphics[width=16cm, height=10.9cm, keepaspectratio]{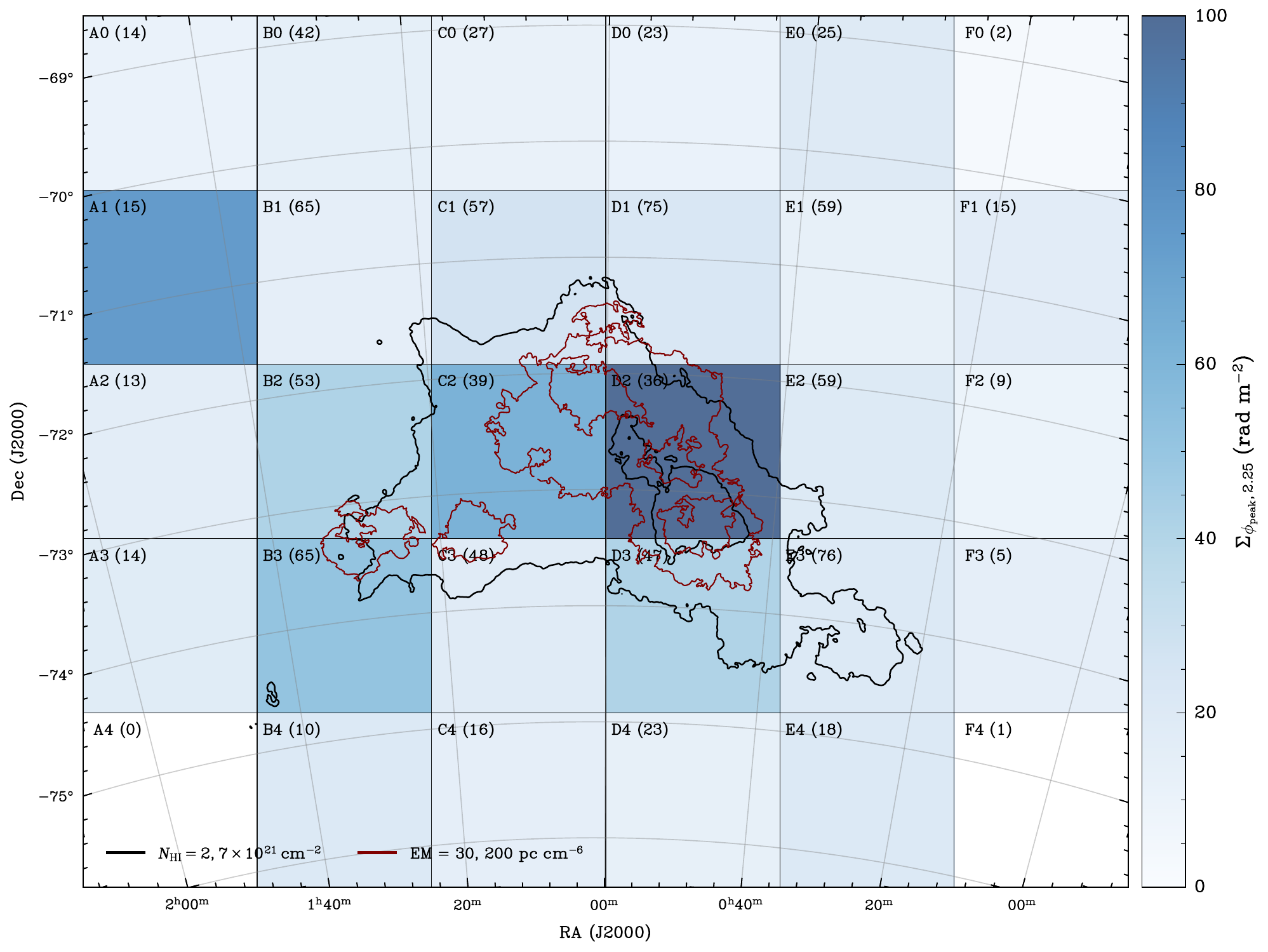}
    \caption{Binned RM grid showing statistical depolarisation and RM dispersion. The upper panel shows the fraction of polarised sources ($\revb{f_{\mathrm{pol}}}$, Equation~\ref{eqn: DP frac ratio}) mapped per $2.25~\mathrm{deg}^2$ across the SMC, with masked regions (indicating no polarised sources) shown in white. Each bin is labelled with a column letter and row number; the number of polarised sources is indicated in parentheses. Black $N_\mathrm{\ion{H}{I}}$ contours at column densities of $2 \times 10^{21}$ and $7 \times 10^{21}$ cm$^{-2}$, derived from ASKAP observations by \citealt{2022.Pingel}, and red EM contours at 20 and 200 $\mathrm{pc}~\mathrm{cm}^{-6}$ from \citealt{2015.Winkler.MCELS} are overlaid, capturing the bulk of the small-scale thermal electron density. Notably, bins C2, D2 and D3 exhibit significantly reduced $\revb{f_{\mathrm{pol}}}$ values compared to the rest of the field. The lower panel displays $\Sigma_{\phi_{\mathrm{peak},2.25}}$ per bin; here, we present the spatial structure of the RM dispersion. With the same contours, bin labelling and polarised source counts as the upper panel.}
    \label{fig: Depaolrisation map per square degree}
\end{figure*}

Figure~\ref{fig: Depaolrisation map per square degree} displays $\revb{f_\mathrm{pol}}$ per $2.25~\mathrm{deg}^2$ bins across the SMC. Each bin is labelled with a column letter and a row number, and the number of polarised sources within each bin is indicated in parentheses. Contours of \ion{H}{I} and EM \rev{for} the SMC, representing the bulk of the neutral and ionised gas, respectively, are overlaid. The error (\revb{$\sigma_{f_{\mathrm{pol}}}$}) remains minimal and uniform across the centre of the field but increases near the field edges where the source density diminishes, with a median value of $\widetilde{\revb{\sigma_{f_{\mathrm{pol},2.25}}}} = 0.054$. Figure \ref{fig: Depaolrisation map per square degree} (upper panel) highlights a notably lower $\revb{f_\mathrm{pol}}$ value in bins encompassing the main \ion{H}{I} body of the SMC, particularly in bins C2 and D2, compared to the surrounding regions, with a median value of $\widetilde{\revb{f_{\mathrm{pol},2.25}}} = 0.26$. This spatial trend is maintained in both 0.25 and 1 square degree binning scales as shown in Figure~\ref{fig: StatDepol 0.5 and 1.5 deg}, demonstrating that regardless of the chosen bin size, the same trend of a statistically lower polarised source density across the main \ion{H}{I} body of the SMC holds. Figure~\ref{fig: Depaolrisation map per square degree} (lower panel) displays $\Sigma_{\phi_\mathrm{peak}}$ per $2.25~\mathrm{deg}^2$ bins across the SMC, showing that bins with low \revb{$\revb{f_\mathrm{pol}}$} coincide with regions of high RM dispersion.

\begin{figure}
    \centering
    \includegraphics[width=\linewidth]{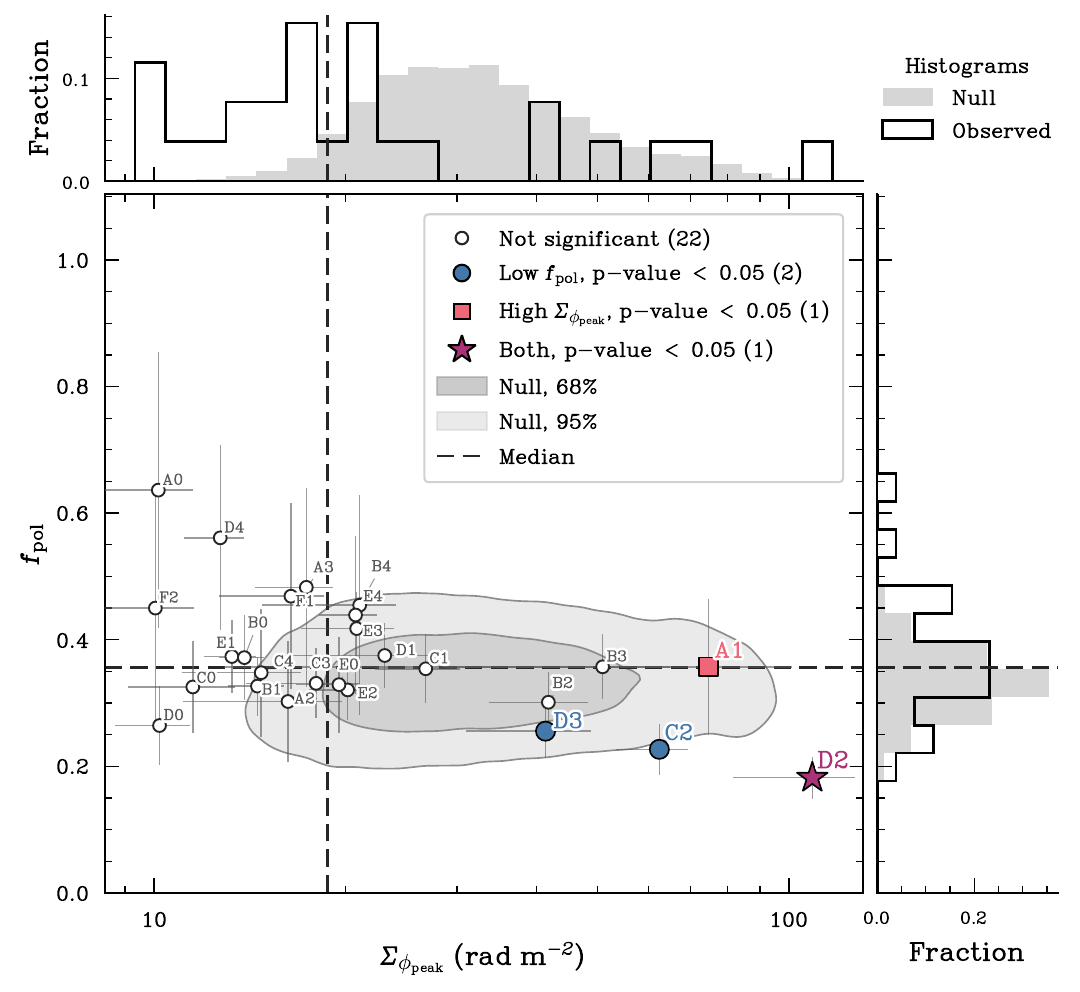}
    \caption{\revb{Fraction of polarised sources, $f_\mathrm{pol}$, against RM dispersion, $\Sigma_{\phi_\mathrm{peak}}$, for the $2.25~\mathrm{deg}^2$ bins containing $N_\mathrm{pol} > 5$ polarised sources. Each bin is labelled by its column letter and row number, following Figure~\ref{fig: Depaolrisation map per square degree}. Error bars show the propagated Poisson uncertainty on $f_\mathrm{pol}$ (Section~\ref{sec 3: Statistical Depolarisation in the SMC}) and the 16th--84th percentile bootstrap uncertainty on $\Sigma_{\phi_\mathrm{peak}}$. Significance is assessed with one-tailed tests against a null distribution obtained by randomly redistributing all source positions across the field ($10{,}000$ iterations). Bins with significantly low $f_\mathrm{pol}$ (blue circles), significantly high $\Sigma_{\phi_\mathrm{peak}}$ (red squares), or both (purple star) have p-value~$< 0.05$; open circles are not significant. The number of bins for each distribution is shown in parentheses in the legend. Grey shaded regions enclose 68 and 95 per cent of the null realisations, pooled over the plotted bins. Dashed lines mark the medians of the observed bins, $\tilde{f}_\mathrm{pol}$ and $\tilde{\Sigma}_{\phi_\mathrm{peak}}$ of Figure~\ref{fig: Depaolrisation map per square degree}. The top and right histograms show the fraction of observed bins (black outline) and of null realisations (grey) in each histogram bin.}}
    \label{fig: D vs sigma RM}
\end{figure}

Figure~\ref{fig: D vs sigma RM} shows the relationship between $\revb{f_\mathrm{pol}}$ and the peak Faraday depth dispersion $\Sigma_{\phi_\mathrm{peak}}$ for spatial bins containing more than five polarised sources, with statistical significance assessed through one-tailed permutation-based spatial randomisation tests ($10{,}000$ iterations). Scatter points show the \revb{$2.25~\mathrm{deg}^2$} bins with their letter-number labels. \revb{Deviations are given in units of $\sigma_\mathrm{null}$, the standard deviation of each bin’s (non-Gaussian) null distribution, alongside empirical one-tailed p-values.} Bin D2 (\revb{purple} star) \revb{deviates significantly from the null in both parameters} (p-value~$< \revb{10^{-4}}$): $\Sigma_{\phi_\mathrm{peak},\mathrm{D2}} = \revb{108.8}~\mathrm{rad~m}^{-2}$ (\revb{$5.0\,\sigma_\mathrm{null}$} above the null \revb{mean}) and $\revb{f_{\mathrm{pol},\mathrm{D2}}} = \revb{0.182}$ (\revb{$3.6\,\sigma_\mathrm{null}$} below \revb{the null mean}). Bins C2 and D3 (blue circles) exhibit \revb{a} significantly low \revb{fraction of polarised sources}, with $\revb{f_{\mathrm{pol},\mathrm{C2}}} = \revb{0.227}$ (\revb{$2.5\,\sigma_\mathrm{null}$} below null, \revb{p-value}~$\revb{= 0.0054}$) and $\revb{f_{\mathrm{pol},\mathrm{D3}}} = \revb{0.255}$ (\revb{$1.8\,\sigma_\mathrm{null}$} below null, \revb{p-value~$= 0.034$}), respectively. Bin A1 (\revb{red} square) shows significantly elevated RM dispersion, with $\Sigma_{\phi_\mathrm{peak},\mathrm{A1}}= 74.7~\rev{\mathrm{rad~m}^{-2}}$ (\revb{$2.8\,\sigma_\mathrm{null}$} above null, \revb{p-value}~$=0.022$). The median \revb{fraction of polarised sources} $\widetilde{\revb{f_\mathrm{pol}}}$ ranges over \revb{$0.343$--$0.391$} across the three binning scales, and the median RM dispersion $\widetilde{\Sigma_{\phi_\mathrm{peak}}}$ over $13.1$--$18.8~\mathrm{rad~m}^{-2}$ \revb{for bins with $N_\mathrm{pol}>5$}. \revb{The median Faraday dispersion of the null distribution is $31.5~\mathrm{rad~m}^{-2}$, with most $2.25~\mathrm{deg}^2$ bins falling below this value (Figure~\ref{fig: D vs sigma RM}), because the Faraday dispersion is spatially concentrated across the main body of the SMC, randomising source positions distributes this structure across the field, resulting in a higher global null distribution.}

\begin{figure*}
    \centering
    \includegraphics[width=\linewidth]{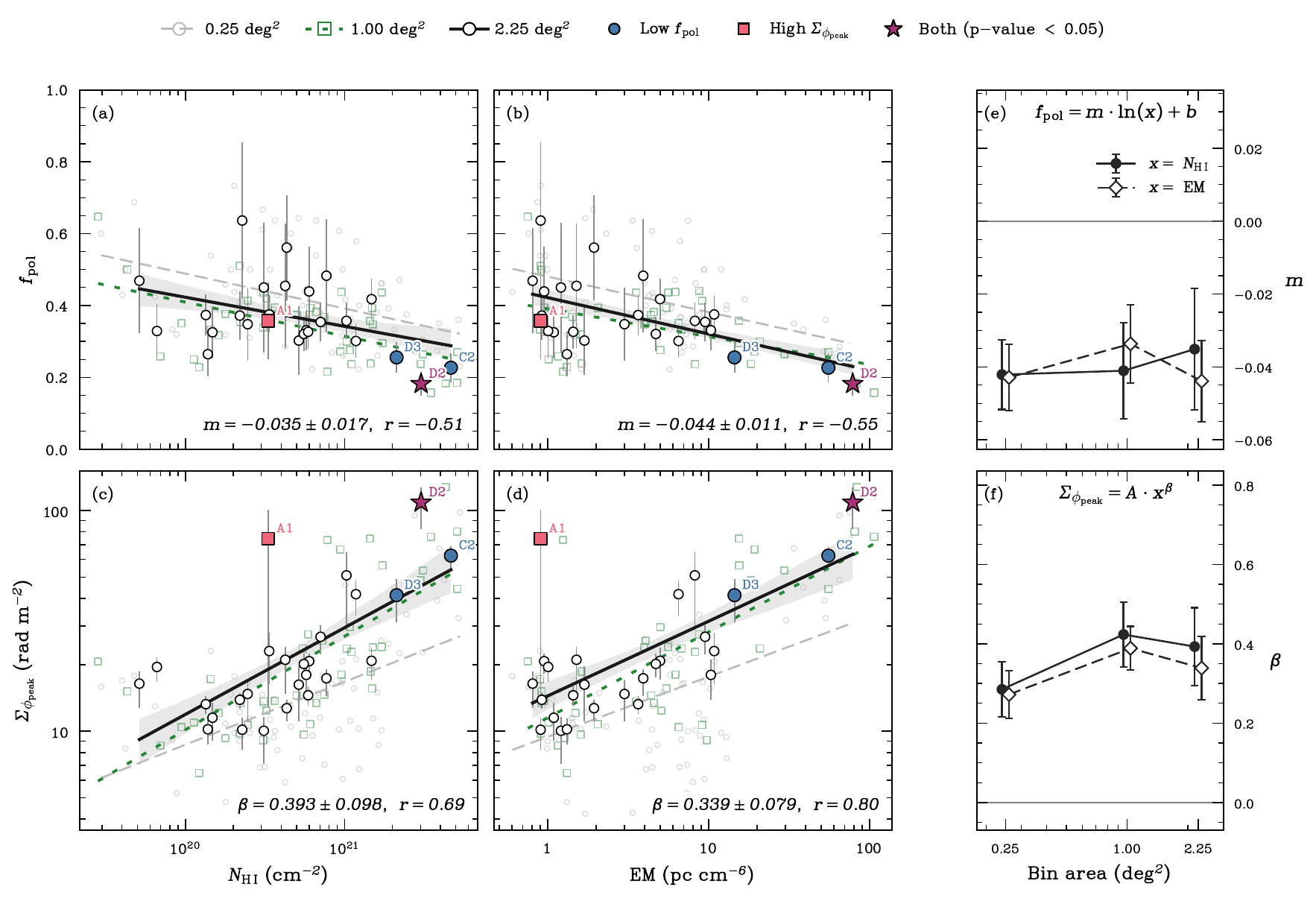}
    \caption{\revb{Fraction of polarised sources, $f_\mathrm{pol}$ (top row), and Faraday dispersion, $\Sigma_{\phi_\mathrm{peak}}$ (bottom row), against the median \ion{H}{I} column density, $N_\mathrm{\ion{H}{I}}$ (panels (a) and (c)), and emission measure, EM (panels (b) and (d)), tracing the neutral and ionised gas, respectively. Only bins with $N_\mathrm{pol} > 5$ are shown. The $2.25~\mathrm{deg}^2$ bins are shown as open circles with error bars, with bins significant in the spatial randomisation test marked as in Figure~\ref{fig: D vs sigma RM}; the $0.25$ and $1.00~\mathrm{deg}^2$ bins are shown as grey circles and green squares respectively. Lines show fits for each bin size: a linear-in-log model, $f_\mathrm{pol} = m 
    \cdot \ln(x) + b$, and a power law, $\Sigma_{\phi_\mathrm{peak}} = A\,x^{\beta}$, fitted in log--log space. Fit uncertainties are from bootstrap resampling ($1000$ iterations), with the shaded region showing the 16th--84th percentile interval of the $2.25~\mathrm{deg}^2$ fit, whose slope and linear Pearson correlation coefficient, $r$, are given in each panel. Panels (e) and (f) show the fitted $m$ and $\beta$ against bin area for both tracers; the horizontal line marks zero.}}
    \label{fig: D and STD-RM with column densities}
\end{figure*}

Figure~\ref{fig: D and STD-RM with column densities} shows the correlations between $\revb{f_\mathrm{pol}}$ and $\Sigma_{\phi_\mathrm{peak}}$ with the median EM and $N_\mathrm{\ion{H}{I}}$ column densities of polarised sources within each bin \revb{(for bins with $N_\mathrm{pol}>5$)}. $\revb{f_\mathrm{pol}}$ decreases with increasing ionised and neutral gas column density, while $\Sigma_{\phi_\mathrm{peak}}$ increases with both tracers, with a stronger dependence on EM, consistent with expectations \citep{2005.Gaensler, 2025MNRAS.541.1106Shah}. For $\revb{f_{\mathrm{pol}}}$, we fit a linear-in-log model, finding $\revb{f_{\mathrm{pol},2.25}} \propto (\revb{-0.035\pm0.016})\ln{(N_\mathrm{\ion{H}{I}})}$ and $\revb{f_{\mathrm{pol},2.25}} \propto (\revb{-0.044\pm0.011})\ln{(\mathrm{EM})}$ at the 2.25 square degree binning scale, with consistent slopes across all binning scales. The linear Pearson correlation coefficients are moderate, ranging from $r=\revb{-0.40}$ to $\revb{-0.54}$ for $N_\mathrm{\ion{H}{I}}$ and from $r=-0.39$ to $\revb{-0.55}$ for EM. For $\Sigma_{\phi_\mathrm{peak}}$, we fit a power-law model, yielding $\Sigma_{\phi_\mathrm{peak,2.25}} \propto N_\mathrm{\ion{H}{I}}^{(\revb{0.393\pm0.093})}$ and $\Sigma_{\phi_\mathrm{peak,2.25}} \propto \mathrm{EM}^{(\revb{0.339\pm0.083})}$ at the 2.25 square degree scale, again with consistent power-law indices across binning scales. The correlations for $\Sigma_{\phi_\mathrm{peak}}$ are systematically stronger than those for $\revb{f_{\mathrm{pol}}}$, with Pearson coefficients ranging from $r = \revb{0.69}$ to $\revb{0.74}$ for $N_\mathrm{\ion{H}{I}}$ and $r = \revb{0.76}$ to $\revb{0.84}$ for EM. We find no discernible correlation between either $\revb{f_{\mathrm{pol}}}$ or $\Sigma_{\phi_\mathrm{peak}}$ and the first or second moments of the \ion{H}{I} emission spectra at any of the three binning scales. 

\revb{Observed statistical depolarisation trends are not dependent on the equal-area square grid adopted here, the same trend is observed in equal-area annuli. Additionally, the fraction of polarised sources is lower within the on-SMC region, with $f_{\mathrm{pol,on}} = 0.227\pm0.021$, compared to $f_{\mathrm{pol,off}} = 0.360\pm0.015$ (see Figure~\ref{fig: histogram of RM on/off SMC}); further on- and off-SMC statistics are given in Table~\ref{tab: Model results and M2 median and mean}.}

\begin{table*}
\resizebox{\textwidth}{!}{%
\begin{tabular}{@{}lccccccc@{}}
\toprule
Catalogue  & Sources Identified  & Mean $\phi_\mathrm{peak}$  & Median $\phi_\mathrm{peak}$  & STD $\phi_\mathrm{peak}$ & Mean $\mathfrak{m}_{2,\mathrm{norm}}$ & Mean $\sigma_{\mathrm{add}}$ & Compact \\ 
& & ($\mathrm{rad~m}^{-2}$) & ($\mathrm{rad~m}^{-2}$) & ($\mathrm{rad~m}^{-2}$) & & & (\%)  \\  
\midrule
\midrule
Polarised & \rev{951} & $10\pm1$ & $15\pm1$ & $39\pm3$ & $0.20\pm0.02$ & \rev{$0.53\pm0.04$} & \rev{76} \\
Off-SMC & \rev{811} & $13\pm1$ & $16\pm1$ & \rev{$29\pm3$} & $0.21\pm0.03$ & \rev{$0.51\pm0.04$} & 75 \\
On-SMC & 140 & $-5\pm6$ & $-1\pm5$ & $74\pm9$ & $0.18\pm0.03$ & $0.66\pm0.19$ & 80 \\ 
\midrule
m1 & \rev{727} & $10\pm1$ & $13\pm1$ & $34\pm2$ & $0.05\pm0.00$ & $0.19\pm0.01$ & 73\\ 
m2 & \rev{87} & $25\pm2$ & \rev{$21\pm1$} & $20\pm5$ & $0.19\pm0.03$ & \rev{$0.68\pm0.07$} & \rev{79}\\
m3 & 42 & $-2\pm12$ & $24\pm4$ & $81\pm19$ & $1.29\pm0.26$ & $2.99\pm0.39$ & 79 \\
m4 & 1 & $22$ & -- & -- & $0.36$ & $1.71$ & 100 \\
m11 & 92 & $5\pm6$ & $7\pm4$ & $53\pm18$ & $0.94\pm0.15$ & $1.85\pm0.29$ & 91 \\
m111 & 2 & $52\pm4$ & $52\pm4$ & $7\pm4$ & $1.26\pm0.47$ & $5.15\pm0.10$ & 50 \\

\bottomrule
\end{tabular}%
}
\caption{Faraday depth statistics of the polarised catalogue, comparing sources on and off the SMC and the results from $QU$-fitting. Columns display the population name, number of identified sources, the mean, median, and standard deviation (STD) of peak Faraday depths, mean $\mathfrak{m}_{2,\mathrm{norm}}$, $\sigma_{\mathrm{add}}$ and percentage of the population that is compact. The uncertainties were determined via bootstrap resampling (1000 iterations). $QU$ fitting identified \rev{224} sources as Faraday complex. Weaknesses of complexity metrics from RM synthesis, $QU$ fitting, $\mathfrak{m}2$ and $\sigma_{\mathrm{add}}$ are discussed in Section \ref{sec: Farady complexity discussion}. On-SMC sources are those with a foreground \ion{H}{I} column density exceeding $2\times 10^{21} ~\mathrm{cm}^{-2}$.}
\label{tab: Model results and M2 median and mean}
\end{table*}

\section{Discussion}\label{sec 5: Discussion}

We now examine the occurrence of depolarisation across the SMC and the potential magneto-ionic structure responsible for it. We consider the complete statistical depolarisation of background sources, then attempt to decouple the magnetic field from the thermal electron density in an effort to identify which fluctuations in the Faraday-rotating medium are driving this depolarisation. Observations of Faraday complexity and its spatial structure are discussed in Appendix~\ref{sec: Farady complexity discussion}.

\subsection{Statistical Depolarisation in the SMC} \label{sec 3: Statistical Depolarisation in the SMC}

The identification of depolarisation through spatial statistics, enabled by the high density of this RM grid, provides a unique means to characterise both the amplitude and spatial scales of magneto-ionic turbulence in the SMC.

If the SMC's RM field is isotropic and follows power-law scaling, then the upper and lower panels of Figure~\ref{fig: Depaolrisation map per square degree} likely sample different length scales of a single turbulent cascade. Under this framework, $\revb{f_{\mathrm{pol}}}$ traces evidence of complete wavelength-dependent beam depolarisation from sub-beam-scale fluctuations (< 6 pc) that produce Faraday dispersion $\sigma_\phi$, while $\Sigma_{\phi_\mathrm{peak}}$ traces larger-scale RM variations.

Elevated $\Sigma_{\phi_\mathrm{peak}}$ across the SMC shows \revb{a strong} correlation with EM and $N_\mathrm{\ion{H}{I}}$ \revb{at all bin sizes (Pearson $r \approx 0.7$ to $0.8$; Figure~\ref{fig: D and STD-RM with column densities})}, indicating large-scale RM variations driven by the turbulent magneto-ionic medium, well beyond the physical scales traced by sub-beam Faraday dispersion. \revb{The fraction of polarised sources, $f_\mathrm{pol}$, is also anticorrelated with both tracers, but more weakly ($r \approx -0.4$ to $-0.55$). Higher SMC gas densities therefore give rise to both enhanced RM dispersion and fewer polarised sources, showing that the SMC is depolarising background sources.} The reduced polarised source densities in bins D2, D3, and C2 indicate strong beam depolarisation, demonstrating that RM fluctuations exist on both large and small scales. \revb{The contrast in correlation strength arises naturally if the electron density is relatively uniform along the LOS on the scale of magnetic fluctuations and acts primarily as a scaling for RM fluctuations (Section~\ref{sec: What is driving depolarisation?}): it sets the amplitude of RM fluctuations on all scales, so the RM dispersion between sightlines closely tracks EM, whereas the sub-beam Faraday dispersion that depolarises sources is much weaker and follows the gas density only loosely, as expected if it is set by the magnetic fluctuations mainly.}

\subsubsection{Bin Features}

Source finding plays an important role in estimating $\revb{f_{\mathrm{pol}}}$, particularly regarding local noise levels and potential continuum contamination from SMC emission. Although we performed source finding on a diffuse-subtracted Stokes $I$ image to minimise extended emission, compact sources \revb{and} non-thermal emitters may still be detected, leading to source count overestimation along the Bar and Wing region, particularly for non-polarised sources. To mitigate this, we applied spectral index cuts (see Section~\ref{sec: 4 Results} and Figure~\ref{fig: Spectral index dsitbruion of the whole catalouge}) and removed known compact \ion{H}{II} regions and planetary nebulae. This established \rev{an approximate} background uniformity in the non-thermal non-polarised distribution, finding mean counts of $12.3 \pm 4.5$ (0.25 deg$^{2}$ bins, $n=97$), $45.6 \pm 14.8$ (1.0 deg$^{2}$ bins, $n=42$), and $122.5 \pm 18.4$ (2.25 deg$^{2}$ interior bins, $n=12$, excluding low-count edge regions). This uniform distribution of non-polarised sources indicates a successful isolation of extragalactic sources, validating the $\revb{f_{\mathrm{pol}}}$ map presented in Figure~\ref{fig: Depaolrisation map per square degree}.

Bin F3, containing five polarised sources, initially appears depolarised but exhibits low $\Sigma_{\phi_\mathrm{peak},2.25}$. Its $\revb{f_{\mathrm{pol}, \mathrm{F3}}}$ value does not persist across spatial scales, suggesting scale-dependent behaviour likely due to beam-edge low-number statistics rather than physical depolarisation. Deeper observations with higher source densities would enable smaller bin sizes without compromising statistical reliability. Furthermore, the complete coverage of the Magellanic clouds by the POSSUM survey would \rev{mitigate} any edge effects \citep{Gaensler_2025}.

\revb{The elevated Faraday dispersion in bin A1 is driven by a single  Faraday simple source, J014316.6$-$713644 with $\phi_\mathrm{peak} = -267.7^{+2.9}_{-2.5}~\mathrm{rad~m}^{-2}$, which lies $\lesssim 13''$ ($\lesssim 110$~kpc, or $\lesssim 0.15\,R_{500}$) from the centre of the background galaxy cluster 1eRASS~J014318.8$-$713652 \citep[photometric $z = 0.965\pm0.008$, $M_{500} \approx 4\times10^{14}\,\mathrm{M}_\odot$;][]{2024A&A...685A.106B}, also detected by Planck as PSZ2~G297.16$-$44.90 \citep{2016A&A...594A..27P}. Its observed Faraday rotation is therefore likely dominated by the intracluster medium; excluding it reduces $\Sigma_{\phi_\mathrm{peak},\mathrm{A1}}$ to $13.7~\mathrm{rad~m}^{-2}$, consistent with the field, and removes its apparent outlier behaviour in the correlations with the neutral and ionised gas tracers (Figure~\ref{fig: D and STD-RM with column densities}).}

\subsection{What is driving depolarisation?} 
\label{sec: What is driving depolarisation?}

The observed spatial statistical depolarisation arises from a turbulent magneto-ionic structure within the beam. This turbulence is characterised by the strength \revb{$(B_r)$} and coherence scale $(l_B)$ of the random magnetic field component, along with the number density fluctuations $(n_e)$ and coherence scale $(l_{n_e})$ of the thermal electrons, each contributing to the observed Faraday depth. The filling fraction $f$ of each component, the fraction of the LOS occupied by coherent structures, may differ between the two. The relative magnitudes of these properties influence which component dominates the depolarisation process \citep{2016.Xu}. For example, in the Milky Way, the $n_e$ fluctuations have a much larger scale ($\approx 300\,{\rm pc}$) compared to the magnetic field fluctuation scale ($\approx 30\,{\rm pc}$) \citep{1995Armstrong, 2025MNRAS.544.2698Dhakal}. To determine the primary driver of the depolarisation, these components must be evaluated separately\footnote{The Faraday-rotating magneto-ionic medium is unlikely to be uniform across the SMC or along the line of sight, and the coupling between the magnetic field and thermal electron density is an important consideration when attempting to disentangle their respective contributions to the observed rotation measure.}.

\revb{In the SMC, the prior observations favour the magnetic field as the primary driver. Although working with a spatially sparser dataset \citep{2022MNRAS.510..260LivingstonSMC}, \citet{SetaEA2023} found $n_e$ approximately constant across the $\approx$160~pc correlation length of the random field, suggesting that on these scales the RM fluctuations are primarily set by the magnetic field rather than by the thermal electron density. This suggests a picture in which the thermal electron fluctuations are smooth along the LOS, concentrated in one or two large ionised regions (large $l_{n_e}$), while the magnetic field is tangled on much smaller scales, with frequent changes in strength and orientation (small $l_B$). In such a medium, $n_e$ cannot vary consistently with the small-scale field, and we therefore attribute the Faraday dispersion to the random magnetic field throughout.}

\begin{figure}
    \centering
    \includegraphics[width=\columnwidth]{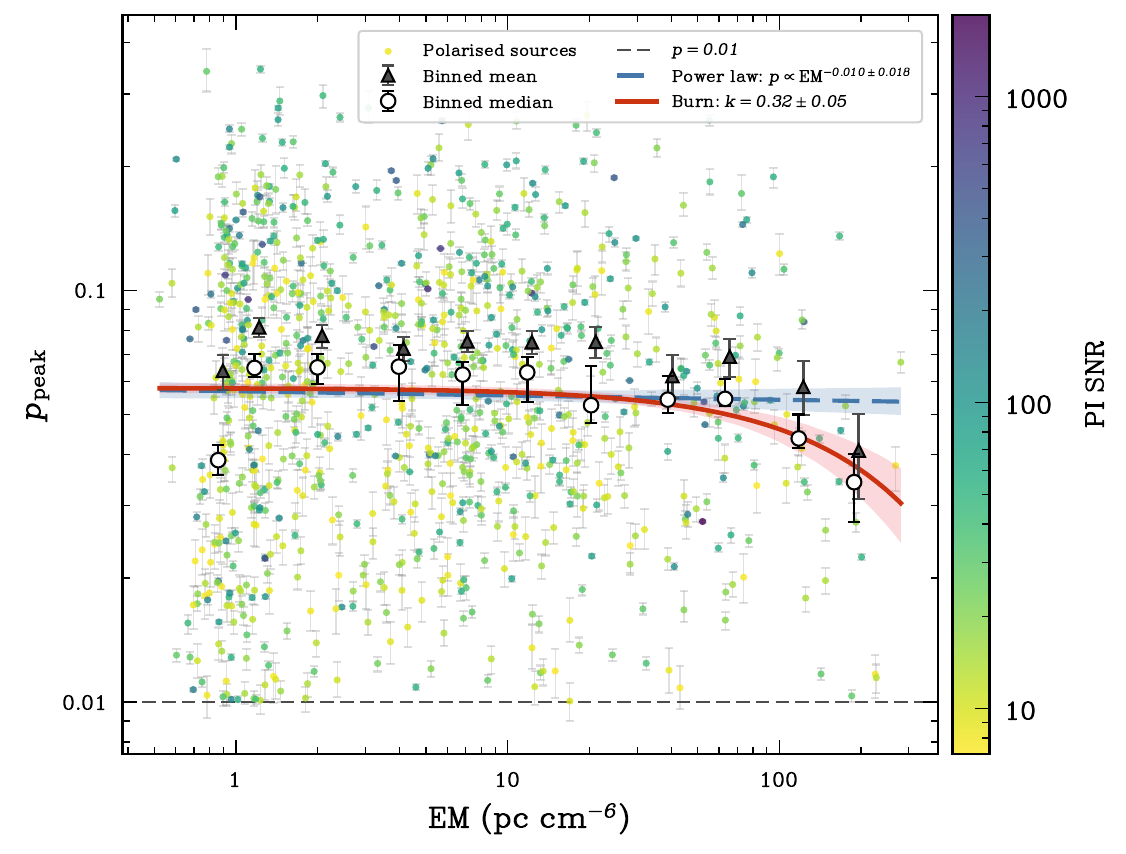}
    \caption{\revb{Peak fractional polarisation against emission measure. The polarised catalogue coloured by their PI SNR (circles), with $1\sigma$ error bars. Sources are binned four per order of magnitude in EM, bin medians (white circles) and means (black triangles) are shown with 68\% intervals. Both lines are maximum-likelihood fits to the individual sources in $\ln p_\mathrm{peak}$, the blue dashed line is a power law, $p \propto \mathrm{EM}^{\,b}$. The red solid line is the Burn model, $p = p_0\,\exp(-2k^2\langle\lambda^4\rangle\,\mathrm{EM})$, i.e.\ $\sigma_\phi = k\sqrt{\mathrm{EM}}$, with $\langle\lambda^4\rangle = 0.0110$~m$^4$ the band average. Shaded regions are the 68\% bootstrap ranges of each fit. The black dashed line marks $p_\mathrm{peak} = 0.01$.}}
    \label{fig: binned fracpol vs ha}
\end{figure}

The relationship between fractional polarisation and EM can be understood through the framework of \cite{2005.Gaensler}, who studied the LMC. Depolarisation arises not from the mean Faraday rotation along a LOS, but from the dispersion in rotation measure $\sigma_\phi$ across the beam. In the \cite{2005.Gaensler} formulation, it is assumed that $n_e$ is uncorrelated with the magnetic field, that \revb{the filling fraction of thermal electrons} $(f_{n_e})$ is small, and that the random component of the magnetic field is constant across the LMC, such that $\sigma_\phi \revb{= k\sqrt{ \mathrm{EM}}}$\revb{, where $k = 0.812\,B_r \sqrt{l_0/3}~(\mathrm{rad~m}^{-2}~\mathrm{pc}^{-1/2}~\mathrm{cm}^{3}$)}.
\revb{This follows from a random walk argument: a sightline of length $L$ crosses $L/l_0$ independent cells of size $l_0$, each adding a Faraday depth of $\approx0.812\,n_e B_\parallel l_0$ with random sign, where $B_\parallel = B_r/\sqrt{3}$, so that $\sigma_\phi = 0.812\,n_e B_r \sqrt{L\,l_0/3} = k\sqrt{\mathrm{EM}}$ with $\mathrm{EM} = n_e^2 L$ (the number of ionised cells and the EM both scale with $f_{n_e}$, so $k$ is unchanged).} Here, EM enters not as the driver of depolarisation but as a scaling factor: regions with higher electron density convert a given level of magnetic field fluctuations into larger RM fluctuations, and hence stronger depolarisation. In Figure~\ref{fig: binned fracpol vs ha}, we show the fractional polarisation of our polarised catalogue as a function of EM. Sources are binned logarithmically with four bins per order of magnitude, and we present the median and mean values for each bin.

\revb{We fit this model to the unbinned data using Burn's law for external Faraday dispersion (beam depolarisation) \citep{1966.Burn, 1991.Tribble}, $p_\mathrm{peak} = p_0\exp(-2\sigma_\phi^2\langle\lambda^4\rangle)$, where $\langle\lambda^4\rangle = 0.0110$~m$^4$ is the POSSUM band average. With $\sigma_\phi = k\sqrt{\mathrm{EM}}$, this becomes $\ln p_\mathrm{peak} = \ln p_0 - 2k^2\langle\lambda^4\rangle\,\mathrm{EM}$, so $\ln p_\mathrm{peak}$ should fall linearly with EM at a rate set by $k$, which measures how much Faraday dispersion the SMC produces at a given EM. We find $p_0 = 0.058\pm0.002$ and $k = 0.32\pm0.05$~rad~m$^{-2}$~pc$^{-1/2}$~cm$^{3}$ (Figure~\ref{fig: binned fracpol vs ha}), with uncertainties determined via bootstrap resampling (1000 iterations). The implied Faraday dispersion is modest: $\sigma_\phi \approx 3$~rad~m$^{-2}$ at $\mathrm{EM} = 100$~pc~cm$^{-6}$, rising to $\approx5.5$~rad~m$^{-2}$ at the highest EMs in our sample ($\approx300$~pc~cm$^{-6}$), where it halves $p_\mathrm{peak}$.
This is no larger than the spread between individual sources, which differ by a factor of $\sim$2 at any EM (0.77 in $\ln p_\mathrm{peak}$), so the SMC's imprint is subtle.}
 
\revb{$QU$-fitting offers an independent route to $k$, measuring $\sigma_\phi$ for individual sources rather than inferring it from the population trend. We use the 102 compact sources best fitted by models m2 and m3 (69 and 33, respectively), which include external Faraday dispersion (Table~\ref{tab: QU-Fitting models fucntions }; the single m4 source is excluded). Assuming that $\sigma_\phi$ for the 88 sources off the SMC arises from intrinsic and Milky Way dispersion, we use them to set a floor of $\sigma_0 = 3.97\pm0.17$~rad~m$^{-2}$. The 14 sources behind the SMC have a higher median, $\sigma_\phi = 4.62^{+0.50}_{-0.27}$~rad~m$^{-2}$. Independent Faraday screens add in quadrature, $\sigma_\phi^2 = \sigma_0^2 + \sigma_{\phi,\mathrm{SMC}}^2$, and fitting the individual sources with $\sigma_{\phi,\mathrm{SMC}} = k\sqrt{\mathrm{EM}}$ yields $k = 0.36^{+0.05}_{-0.06}~\mathrm{rad~m}^{-2}~\mathrm{pc}^{-1/2}~\mathrm{cm}^{3}$, consistent with the Burn's law fit (Figure~\ref{fig: binned fracpol vs ha}). Subtracting the floor in quadrature from the median gives an SMC contribution of $\sigma_{\phi,\mathrm{SMC}} \approx 2.4$~rad~m$^{-2}$ along these sightlines, whose median EM is 63~pc~cm$^{-6}$, equivalent to $k = \sigma_{\phi,\mathrm{SMC}}/\sqrt{\mathrm{EM}} \approx 0.3~\mathrm{rad~m}^{-2}~\mathrm{pc}^{-1/2}~\mathrm{cm}^{3}$.}

\revb{Our measured value of $k$ is a factor of $\sim6$ lower than that found in the LMC \citep[$k \approx 1.8$~rad~m$^{-2}$~pc$^{-1/2}$~cm$^{3}$;][]{2005.Gaensler}, indicating substantially weaker Faraday dispersion in the SMC at a given EM. For $\mathrm{EM} = 100$~pc~cm$^{-6}$, the LMC value of $k$ implies $\sigma_\phi \approx 18$~rad~m$^{-2}$, sufficient to completely depolarise background sources across our observing band, compared with $\sigma_\phi \approx 3$~rad~m$^{-2}$ for our measured $k$. Given that $k \propto B_r\sqrt{l_0}$, this difference is consistent with a weaker random magnetic field, a smaller turbulent cell size, or both, in the SMC. However, \citet{2005.Gaensler} measured $k$ at 1.4~GHz, where Faraday dispersions of up to $\sim$20~rad~m$^{-2}$ remain detectable, using sightlines largely through the LMC itself, whereas at our lower frequencies sources with $\sigma_\phi \gtrsim 12$~rad~m$^{-2}$ are depolarised below the detection limit, and our sample extends well beyond the main body of the SMC (Figure~\ref{fig: RMs from QU fitting in the sky}). The spatially coherent statistical depolarisation observed across the SMC (Figure~\ref{fig: Depaolrisation map per square degree}) indicates that the magneto-ionic turbulence is not uniform across the SMC; sightlines through the most turbulent gas are not observed (Figure~\ref{fig: D and STD-RM with column densities}), biasing our estimate of $k$ towards less turbulent environments. Our measured $k$ therefore characterises the less turbulent gas through which polarised background sources survive, and is effectively a lower limit.}

\revb{Translating $k$ into a field strength requires knowledge of the cell size. For $k = 0.32$, $B_r = 1.0\pm0.2\,(l_0/0.5~\mathrm{pc})^{-1/2}~\mu$G, normalised to the 0.5~pc cells that \citet{2005.Gaensler} adopted for the LMC: larger cells give fewer but larger steps in the random walk, so the same dispersion requires a weaker field. Depolarisation requires many cells within each $\sim 6~\mathrm{pc}$ beam, so $l_0 \lesssim 6$~pc and $B_r \gtrsim 0.3~\mu$G. In turbulence with a continuous range of scales, this bound becomes an estimate: the Faraday dispersion within a beam is dominated by the largest eddies that still fit inside it, so the effective cell size is close to the beam itself and $B_r \approx 0.3~\mu$G on $\sim 6~\mathrm{pc}$ scales. This estimate assumes that H$\alpha$ traces the Faraday-rotating gas, and becomes an upper limit if fluctuations in electron density also contribute to $\sigma_\phi$.}

\revb{\citet{SetaEA2023} also measured the spectrum of the turbulent field on larger scales, using RM structure functions. They find a correlation length of $l_B = 160\pm21$~pc, a random-field strength of $4.7\pm0.6~\mu$G, POS RM fluctuations between independent sightlines reaching $\Sigma_\mathrm{RM} \approx 117$~rad~m$^{-2}$, and a magnetic power spectrum of slope $-1.3\pm0.4$ below $l_B$. Because the SMC is far deeper along the LOS \citep[$L \sim 9$~kpc;][]{2019.Smart} than $l_B$, every sightline crosses at least $L/l_B \approx 56$ independent cells, so RM differences accumulate as a random walk; with $n_e$ nearly constant, the RM structure function then rises as $r^{1.3}$ below $l_B$, with the same slope as the magnetic power spectrum. Their sightlines have a mean H$\alpha$ intensity of $\approx29~\mathrm{R}$, corresponding to $\mathrm{EM} \approx 110$~pc~cm$^{-6}$ for $T_\mathrm{e} = 1.4\times10^4$~K \citep{1977ApJ...216..433Reynolds}. If the field keeps this spectrum down to smaller scales, the RM dispersion within a beam of size $l_\mathrm{beam} < l_B$, which goes as the square root of the structure function, is $\sigma_\phi \approx \Sigma_\mathrm{RM}\,(l_\mathrm{beam}/l_B)^{0.65}$. For our 6~pc beam, this predicts $\sigma_\phi \approx 117\,(6/160)^{0.65} \approx 14$~rad~m$^{-2}$, whereas at $\mathrm{EM} \approx 110$~pc~cm$^{-6}$ our Burn's law fit gives $\sigma_\phi = k\sqrt{\mathrm{EM}} \approx 3.4$~rad~m$^{-2}$, four times less.
The same holds for the field strength: their $4.7\pm0.6~\mu$G assumes a filling factor of 0.4, equivalent to $B_r \approx 1.9~\mu$G at 160~pc in our formulation, in which the filling factor cancels; with $\delta B \propto l^{(1.3-1)/2} = l^{0.15}$, this falls only to $\approx1.2~\mu$G at 6~pc, against the $\approx0.3~\mu$G we infer. Reconciling the two requires a steeper magnetic power spectrum between the beam and 160~pc, with a slope of $\approx -2$, steeper than both their fitted slope and Kolmogorov ($-5/3$), but within $\sim$2$\sigma$ of their measurement. Our depolarisation therefore probes the small-scale end of the turbulence they measured, and indicates less Faraday-rotating power on parsec scales than a single power law from 160~pc implies.}

However, these interpretations are sensitive to the three-dimensional structure of the SMC, which remains uncertain \citep{1987.Mathewson, 2017.Ripepi, 2024.Murray}. The SMC likely comprises two components separated by $\sim5$ kpc, implying that the characteristic length scales driving depolarisation may vary spatially. A more complete understanding of the SMC's geometry is needed to robustly disentangle the respective contributions of $n_e$ and \revb{$B_r$}, \revb{a goal of future work.}

Nevertheless, the overall picture of magnetic-field-driven depolarisation is consistent with other extragalactic studies, which through modelling and multi-wavelength analysis typically attribute depolarisation to a turbulent magnetic field \citep{2011.Fletcherrql, 2017.Basu, 2020.Kierdorf}. This is further supported by Galactic observations indicating that the thermal electron distribution is more uniform across scales, while the magnetic field structure is more influential \citep{1991pihp.book..159Marsch, 1996.Minter, 2008.Haverkorn, 2025MNRAS.544.2698Dhakal}. High angular resolution observations of $n_e$, such as those anticipated from the Sloan Digital Sky Survey V Local Volume Mapper \citep{2024AJ....168..198Drory}, can provide the spatial detail needed to constrain the role of thermal electron fluctuations in driving depolarisation.

\section{Conclusion}\label{sec: 7 summary}
Using the Polarisation Sky Survey of the Universe's Magnetism (POSSUM) Pilot II survey on the Australian SKA Pathfinder (ASKAP), we present a high-density rotation measure (RM) grid of \rev{951} broadband polarised background radio sources ($\sim24$ RMs per square degree) to probe the magneto-ionic structure of the Small Magellanic Cloud (SMC). Utilising the radio polarimetric analysis algorithms of RM-synthesis and Stokes $QU$-fitting, with each source fit using six distinct Faraday models, this work presents, to our knowledge, the first large-scale application of $QU$-fitting to such a dense RM grid across this broad a bandwidth.

We characterised the SMC's magneto-ionic properties using two spatial measures - the fraction of polarised sources and RM dispersion between sources - complemented  by RM dispersion towards source where observed. We identified spatially coherent depolarisation ($\revb{f_{\mathrm{pol}}}$) signatures across the SMC that correlated with enhanced plane-of-sky RM fluctuations ($\Sigma_{\phi_\mathrm{peak}}$) (Figure~\ref{fig: Depaolrisation map per square degree}), revealed through a statistical comparison of polarised and non-polarised non-thermal source densities on a per-square-degree basis. The absence of polarised sources across the main body of the SMC is statistically significant, with depolarisation strengthening in regions of elevated RM dispersion. Both quantities are correlated with emission measure and neutral hydrogen column density, and persistent across a range of binning scales. These trends demonstrate that the magneto-ionic medium of the SMC drives both enhanced RM dispersion and depolarisation of background sources. \revb{Population-level modelling of beam depolarisation with emission measure indicates that the SMC produces $\sim6$ times less Faraday dispersion than the Large Magellanic Cloud at the same emission measure, although differences in observing frequency and sightline distribution likely contribute to the discrepancy. Additionally, complete depolarisation of sources behind the most turbulent gas of the SMC biases our fit towards less turbulent sightlines.}

Faraday complexity was detected in $\sim24$\% of the RM catalogue; however, no coherent spatial structure was observed in the complexity distribution. We attribute this complexity primarily to internal source structure or an extended halo surrounding the SMC, rather than to the magneto-ionic medium of the SMC main body. Standard RM synthesis-based complexity metrics routinely used in the literature ($\mathfrak{m_{2,norm}}$ and $\sigma_\mathrm{add}$) failed to recover the full population of complex sources identified through $QU$-fitting, with their classification accuracy strongly dependent on polarised \revb{intensity} signal-to-noise ratio (see Appendix~\ref{sec: Farady complexity discussion}).

Taken together, these results provide compelling evidence that the SMC depolarises background radio sources through a turbulent magneto-ionic medium on scales smaller than the beam ($\leq 6~\mathrm{pc}$). \revb{This depolarisation is most likely driven by small-scale fluctuations in the magnetic field, which on these scales must have a random field strength of $B_r \gtrsim 0.3~\mu$G.} This work demonstrates the capability of dense RM grids to resolve the detailed line-of-sight magnetic field structure of the interstellar medium.

\section*{Acknowledgements}

J.~M.~P thanks Kathleen A. Barger, Sui Ann Mao\rev{, and Rainer Beck} for their thoughts and comments on this work, and \rev{Luka Mijnarends and Eric Muller for their support}. \revb{We thank the anonymous reviewer, whose comments have improved the quality and rigour of this paper.} This research was partially funded by the Australian Government through an Australian Research Council Australian Laureate Fellowship (project number FL210100039 awarded to N.~M.~Mc-G.) and a Discovery Project (DP220101558). J.~M.~P acknowledges the German Academic Exchange Service (DAAD). Y.~K.~M.\ is partially supported by the Federal Ministry of Education and Research (BMBF) in Germany via the German Academic Exchange Service (DAAD) Project-based Personnel Exchange Programme (PPP). A.~S. is supported by the Australian Research Council through the Discovery Early Career Researcher Award (DECRA) Fellowship (project DE250100003) funded by the Australian Government and the Australia-Germany Joint Research Cooperation Scheme of Universities Australia (UA--DAAD, 2025--2026). S.~L.~J acknowledges support from the UK Research and Innovation (UKRI) Frontiers Research Grant [EP/X026639/1]. J.~D.~L is funded under the Mapping Magnetic Fields with INterferometry Down to Event hoRizon Scales (M2FINDERS) project. The M2FINDERS project has received funding from the European Research Council (ERC) under the European Union’s Horizon 2020 research and innovation programme (grant agreement No 101018682). S.~P.~O acknowledges support from the Comunidad de Madrid Atracción de Talento program via grant 2022-T1/TIC-23797, and grant PID2023-146372OB-I00 funded by MICIU/AEI/10.13039/501100011033 and by ERDF, EU. 

This scientific work uses data obtained from Inyarrimanha Ilgari Bundara, the CSIRO Murchison Radio-astronomy Observatory. We acknowledge the Wajarri Yamaji People as the Traditional Owners and native title holders of the Observatory site. CSIRO’s ASKAP radio telescope is part of the Australia Telescope National Facility (https://ror.org/05qajvd42). Operation of ASKAP is funded by the Australian Government with support from the National Collaborative Research Infrastructure Strategy. ASKAP uses the resources of the Pawsey Supercomputing Research Centre. Establishment of ASKAP, Inyarrimanha Ilgari Bundara, the CSIRO Murchison Radio-astronomy Observatory and the Pawsey Supercomputing Research Centre are initiatives of the Australian Government, with support from the Government of Western Australia and the Science and Industry Endowment Fund.

\section*{Data Availability}

This paper makes use of raw data products that are publicly available through the CSIRO ASKAP Science Data Archive (CASDA; \cite{2017.Chapman, Huynh2020}). The processed POSSUM data used in this work is subject to the guidelines of the POSSUM survey and will be shared upon reasonable request to the corresponding author.


\bibliographystyle{mnras}
\bibliography{references}


\appendix

\section{Supplementary Material}\label{sec: Appendix sub figs}

\subsection{Statistical Depolarisation at smaller scales}
\begin{figure*}
    \centering
    \includegraphics[width=0.48\textwidth]{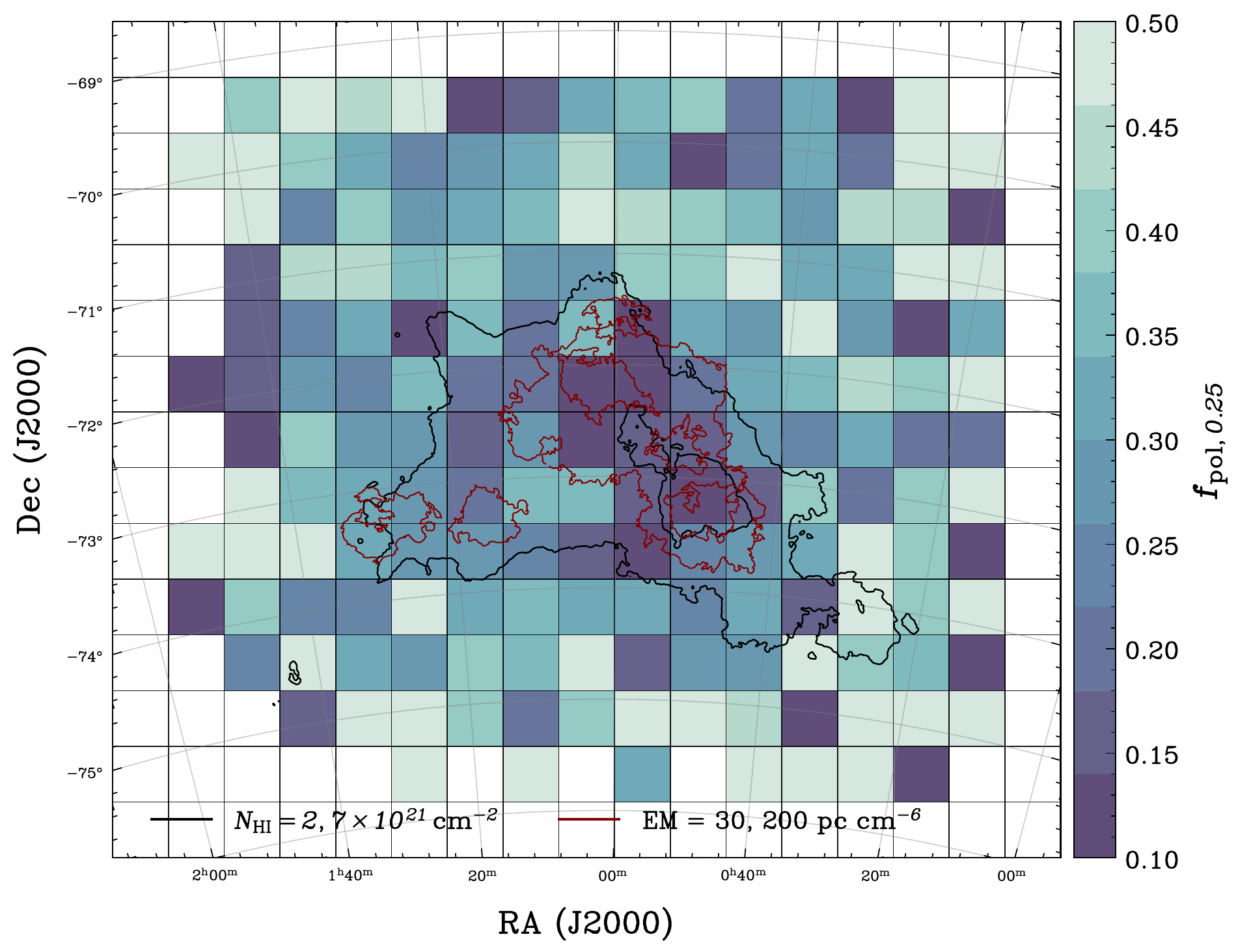}
    \hfill
    \includegraphics[width=0.48\textwidth]{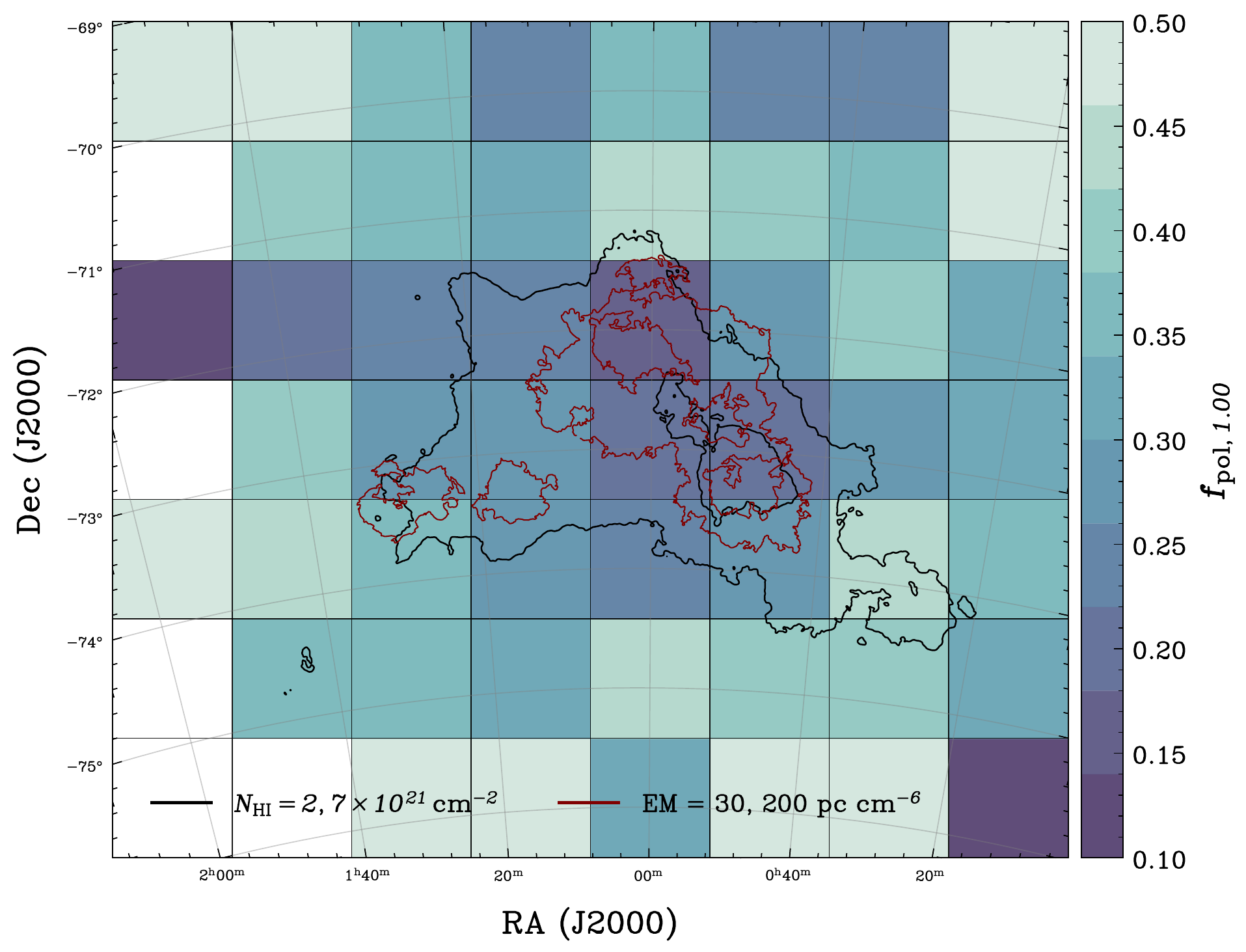}
    \\[2mm]
    \includegraphics[width=0.48\textwidth]{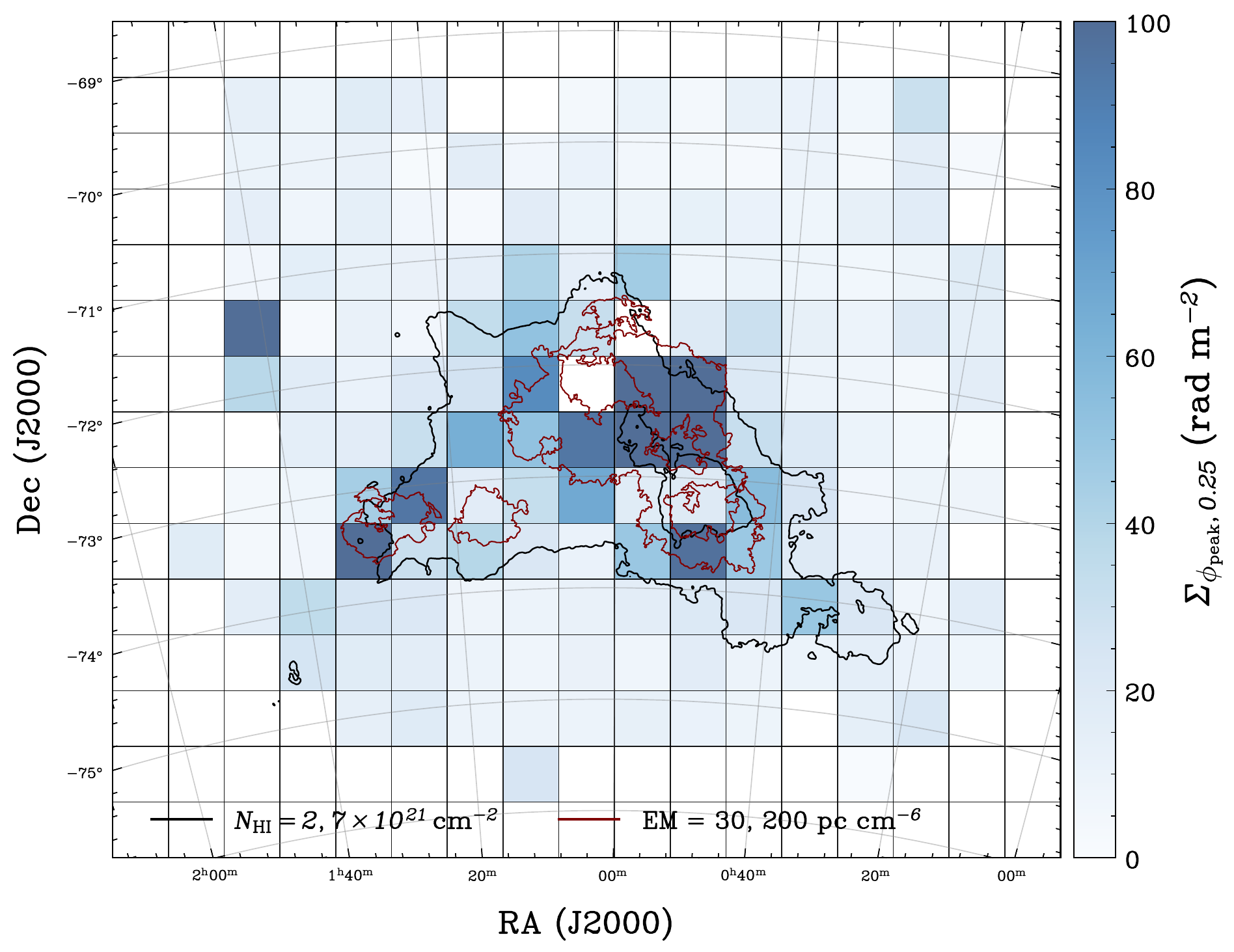}
    \hfill
    \includegraphics[width=0.48\textwidth]{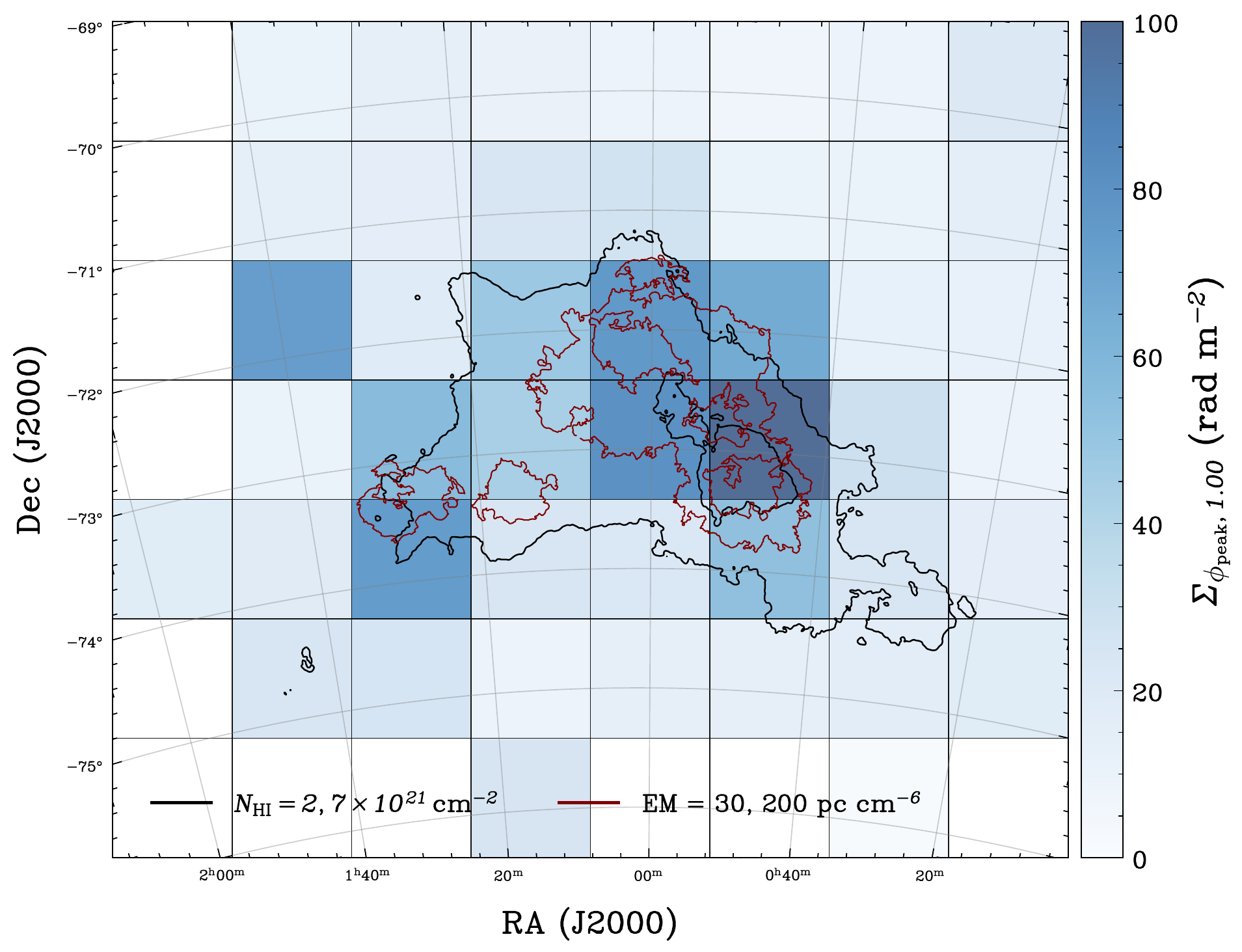}
    \caption{Fraction of polarised sources, $f_\mathrm{pol}$ (top row), and peak Faraday depth dispersion, $\Sigma_{\phi_\mathrm{peak}}$ (bottom row), for $0.25~\mathrm{deg}^2$ (left) and $1.00~\mathrm{deg}^2$ (right) bins across the SMC. Black $N_{\mathrm{HI}}$ contours at column densities of $2 \times 10^{21}$ and $7 \times 10^{21}$~cm$^{-2}$, derived from ASKAP observations by \citealt{2022.Pingel}, and red EM contours at 30 and 200~$\mathrm{pc}~\mathrm{cm}^{-6}$ from \citealt{2015.Winkler.MCELS} are overlaid, capturing the bulk of the small-scale thermal electron density. Both quantities show the same spatial trends as the $2.25~\mathrm{deg}^2$ maps in Figure~\ref{fig: Depaolrisation map per square degree}, demonstrating that the trends persist across binning scales.}
    \label{fig: StatDepol 0.5 and 1.5 deg}
\end{figure*}

Figure~\ref{fig: StatDepol 0.5 and 1.5 deg} present $\revb{f_{\mathrm{pol}}}$ at smaller binning scales compared to those in Figure~\ref{fig: Depaolrisation map per square degree}, at 0.25 (left) and 1.0 (right) square degrees respectively. These demonstrate that the spatial statistical depolarisation trend is preserved across binning scales, in addition to Figure~\ref{fig: D and STD-RM with column densities}.

\subsection{RM Jitter}

RM jitter is the additional scatter in measured rotation measures produced by unresolved Faraday rotation fluctuations in turbulent plasma local to the source on sub-beam scales, and is most evident statistically in populations of sources with low fractional polarisation \citep{2025.Stil}. Following the approach of \cite{2024.Vanderwoude}, who identified RM jitter potential signatures by comparing the interquartile range (IQR) of RM distributions between weakly polarised ($<3\%$) fractional polarisation) and strongly polarised ($\geq3\%$) source populations, we find no significant difference between the two samples in our field: the IQR for low- and high-polarisation sources is $30.96~\mathrm{rad
~m^{-2}}$ and $29.66~\mathrm{rad~m^{-2}}$, respectively, in contrast to the $\sim2.5
\times$ difference reported by \cite{2024.Vanderwoude}. This similarity persists when restricting to off-SMC sources ($28.01$ vs $26.46~\mathrm{rad~m^{-2}}$). Instead, the dominant source of RM scatter is the large-scale Faraday structure of the SMC itself, as evidenced by the substantially higher IQR for on-SMC sources ($54.57~\mathrm{rad~m^{-2}}$) compared to off-SMC sources ($26.99~\mathrm{rad~m^{-2}}$). We conclude that RM jitter does not significantly contribute to the observed RM dispersion in this field, and does not impact the conclusion that the SMC is depolarising background sources.

\subsection{Anomalous Sources}\label{sec: Unique sources}

In this section we provide a brief discussion of the spectro-polarimetrically anomalous sources within the catalogue. We define these as sources that were successfully fitted by Stokes $QU$-fitting using a Faraday-simple model (m1), but which possess a $\sigma_{\mathrm{add}} > 1$. A high $\sigma_{\mathrm{add}}$ indicates that the observed fractional Stokes $q$ and $u$ spectra deviate significantly from the behaviour expected for a single Faraday-thin component, suggesting Faraday complexity. There are nine such sources in the field, which we list in Table~\ref{tab: sigma_add_sources}.

Several physical and observational effects may give rise to such anomalous behaviour. At the angular resolution of these observations, the most likely cause is unresolved internal structure within the emitting component or Faraday-active plasma local to the source. The $QU$-fitting models applied in this work examines multi-component Faraday structure and external Faraday dispersion; however, it does not include models that produce repolarisation, the phenomenon in which fractional polarisation increases with wavelength, sometimes referred to as inverse depolarisation \citep[e.g.][]{2002.Homan, 2012AJ....144..105H, 2012.Homan, 2014.Farnes}. A variety of physical scenarios have been proposed to produce repolarisation. For example,  \cite{2012.Homan} proposed a model in which structural inhomogeneities in a jet magnetic field and internal Faraday rotation give rise to this effect. In this model, differences in the magnetic field structure between the near and far sides of the jet lead to partial cancellation of polarisation along the line of sight, resulting in low net polarisation at short wavelengths. At longer wavelengths, internal Faraday rotation, which grows with wavelength squared, acts to align the polarisation contributions from the two sides of the jet, reducing the cancellation and producing a net increase in fractional polarisation. One might also expect repolarisation to arise naturally in unresolved, optically thick core regions where different wavelengths probe different physical depths within the jet \citep{1981ApJ...243..700Konigl}.

\begin{figure*}
    \centering
    \includegraphics[width=\linewidth]{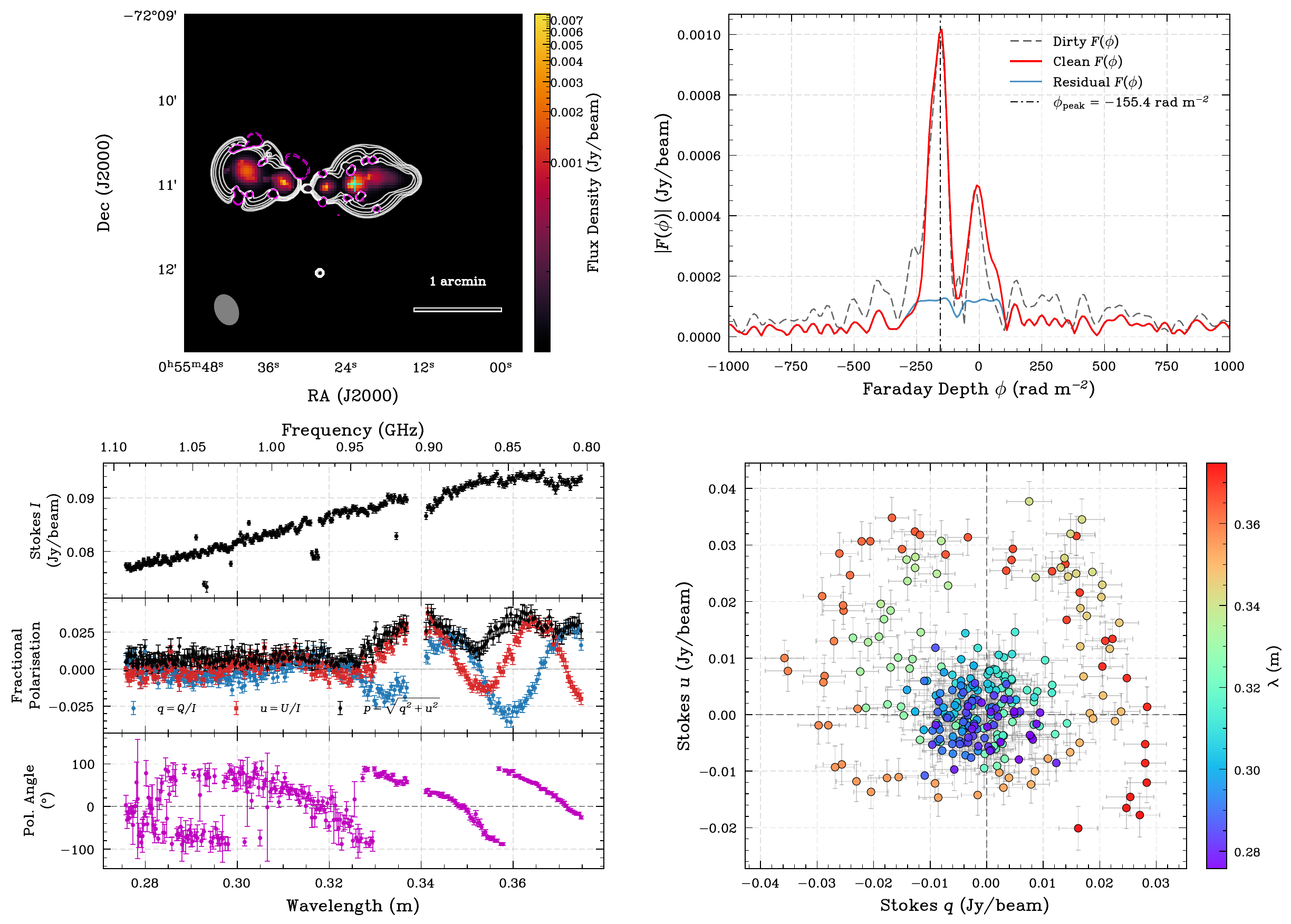}
    \caption{Anomalous source J005523.4-721056, exhibiting repolarisation features. Top left: Stokes $I$ intensity from the MFS image. The synthesised beam is shown in the lower left corner, with a $21''$ major axis. White contours indicate positive emission at levels of 3, 5, 10, 20, 50, and 100$\sigma$, while magenta dashed contours trace negative emission at $-3\sigma$ and $-5\sigma$, where $\sigma = 0.169~\mu\mathrm{Jy/beam}$. Top right: Faraday spectrum \rev{$F(\phi)$} showing the dirty (dashed black), clean (solid red), and residuals (solid blue). The vertical dot-dashed line marks the peak Faraday depth at $\phi_{\mathrm{peak}} = \revb{-155.4}~\mathrm{rad~m^{-2}}$. Bottom left: Spectral properties as a function of wavelength: Stokes $I$ intensity (top), fractional Stokes $q$ and $u$ with fractional polarisation $p$ (middle), and polarisation angle $\psi$ (bottom). A secondary frequency axis is shown along the top. Bottom right: Fractional Stokes $q$ versus $u$, colour-coded by wavelength.}
    \label{fig: repolarisation}
\end{figure*}

We show one such repolarising source, J005523.4$-$721056, as an example in Figure~\ref{fig: repolarisation}. The spectral structure of this source exhibits a clear repolarisation signature at longer wavelengths. Our $QU$-fitting procedure initially excluded this feature and fits only the longer-wavelength data, which correspond to the peak Faraday depth recovered by RM synthesis, yielding $\phi = \revb{-155.4}~\mathrm{rad~m^{-2}}$. This source was also identified by \citet{2022MNRAS.510..260LivingstonSMC}, who reported $\mathrm{RM} = -157.7 \pm 7.6~\mathrm{rad~m^{-2}}$ from $1.4-3.0~\mathrm{GHz}$ (see their Figure~46 in supplementary material), consistent with our measurement. We note that the repolarisation characteristics of this source bear a superficial resemblance to those sometimes seen pulsars; however, the MFS image reveals that the emission is associated with a bright component of a radio-loud AGN with a morphology consistent with an FR~II galaxy.

\begin{table*}
\centering
\resizebox{\textwidth}{!}{
\begin{tabular}{ccccccccccccc}
\hline
Source Name & RA & Dec & $\phi_\mathrm{peak}$ & $p$ & $\chi_0$ & SNR$_\mathrm{PI}$ & SNR$_\mathrm{I}$ & $\alpha$ & $\mathfrak{m}_{2,\mathrm{norm}}$ & $\sigma_{\rm add}$ & $\ln\mathcal{E}$ & Compact \\
 & (deg) & (deg) & (rad m$^{-2}$) & (\%) & (deg) &  &  &  &  &  &  &  \\
\hline
\hline
J013243.9-734413 & 23.1829 & -73.7372 & $-20.83 \pm 0.45$ & 2.39 & $138.3 \pm 2.5$ & 65.9 & 1509.0 & $-0.83 \pm 0.01$ & 0.40 & 1.46 & 1500.6 & Y \\
J010255.7-743536 & 15.7323 & -74.5936 & $-0.86 \pm 0.54$ & 4.23 & $55.9 \pm 3.0$ & 54.6 & 824.6 & $-0.71 \pm 0.01$ & 0.19 & 1.12 & 1196.4 & Y \\
J004936.2-741547 & 12.4009 & -74.2631 & $-26.38 \pm 0.12$ & 14.05 & $6.7 \pm 0.7$ & 242.6 & 1198.5 & $-0.84 \pm 0.01$ & 0.21 & 2.24 & 865.4 & Y \\
J002008.4-751225 & 5.0349 & -75.2072 & $1.75 \pm 1.12$ & 4.24 & $143.6 \pm 6.2$ & 27.0 & 375.9 & $-0.42 \pm 0.02$ & 0.33 & 1.19 & 791.0 & Y \\
J014235.7-723001 & 25.6487 & -72.5004 & $-4.32 \pm 0.28$ & 6.00 & $79.1 \pm 1.6$ & 107.2 & 1576.0 & $-0.97 \pm 0.01$ & 0.16 & 1.09 & 1368.0 & Y \\
J005533.6-723126 & 13.8902 & -72.5241 & $-280
.66 \pm 1.19$ & 1.63 & $63.0 \pm 6.5$ & 25.1 & 709.3 & $-0.69 \pm 0.01$ & 0.51 & 1.27 & 1243.6 & Y \\
J005523.4-721056 & 13.8474 & -72.1823 & $-155.40 \pm 0.56$ & 1.15 & $160.0 \pm 3.0$ & 54.1 & 1921.3 & $-0.72 \pm 0.00$ & 1.37 & 3.63 & 240.1 & Y \\
J004413.8-724302 & 11.0577 & -72.7173 & $-27.20 \pm 0.41$ & 2.74 & $136.0 \pm 2.1$ & 75.8 & 1456.8 & $-0.30 \pm 0.01$ & 0.29 & 1.02 & 1654.6 & Y \\
J003401.1-702626 & 8.5045 & -70.4407 & $18.51 \pm 0.13$ & 5.03 & $106.8 \pm 0.7$ & 234.1 & 2014.3 & $-0.72 \pm 0.00$ & 0.27 & 1.62 & 1723.7 & Y \\
\hline
\end{tabular}}
\caption{Anomalous sources best fit by $QU$-fitting with Model 1 as Faraday simple but with $\sigma_{\rm add} > 1$. These sources are not considered in the analysis of the main text. This table identities the nine such sources in this catalogue with their source Name, Right Ascension, Declination, peak Faraday depth from RM synthesis, fractional polarisation, de-rotated polarisation angle, signal-to-noise ratio in polarised intensity, signal-to-noise ratio in Stokes $I$, normalised m2, $\sigma_\mathrm{add}$ (see Section~\ref{sec: rm synth Faraday Complexity Metrics}), log Bayesian evidence and source compactness a source is consider compact if it speak flux is within 10\% of its total flux from the source finder \texttt{pyBDSF} (compact: Y, extended: N) (see Section~\ref{sec: 2 Observations}).}
\label{tab: sigma_add_sources}
\end{table*}

\subsection{Interpretation of Faraday complexity}\label{sec: Farady complexity discussion}

\subsubsection{Astrophysical Faraday Complexity}

The observed Faraday complexity in this study does not appear to arise primarily from the SMC main body, but rather from effects intrinsic to the background sources, or possibly from some extended structure around the SMC. The spherical Voronoi tessellation in Figure~\ref{fig: Voronoi Faraday complexity}, reveals no distinct spatial structure in the distribution of Faraday complex sources or of those exhibiting Faraday dispersion. We find no correlation between Faraday complexity and \ion{H}{I} column density or H$\alpha$ emission.

Previous studies \citep[e.g.,][]{2011.Law, 2014.Rudnick} have demonstrated that multi polarisation component sources can arise from beam effects or limitations in source-finding algorithms used in polarimetric surveys. In particular, consider the case of Fanaroff-Riley Class II (FR II) sources, for which the polarised emission from the bipolar jets can individually experience dissimilar Faraday rotation from their local magneto-ionic medium \citep{1988Natur.331..147G, 1988Natur.331..149L, 2011.Law}. If such FR II sources are (at least partially) spatially resolved, our extraction method that identifies the peak brightness at the pixel level should be able to identify the two lobes as separate spatial components, rather than combining them into a single component. However, if the FR II source is unresolved, the two spatial components can be combined to produce a multi-component Faraday spectrum that is related to the physical structures within/surrounding the source. As shown in Table~\ref{tab: Model results and M2 median and mean}, 91\% of m11 sources and 79\% of m3 sources are considered compact, the single m4 sources, along with one m111 are also compact. The investigation of m111 and m4 sources is beyond the scope of this work but they are likely unrelated to the SMC. The 24\% Faraday complex fraction of this sample is most likely a reflection of the observational parameters, as Faraday complexity is known to depend on frequency coverage and Stokes $I$ signal-to-noise ratio \citep{2011.Law, 2017.O'Sullivan, 2026MNRAS.547ag394O}.

\begin{figure}
    \centering
    \includegraphics[width=\linewidth]{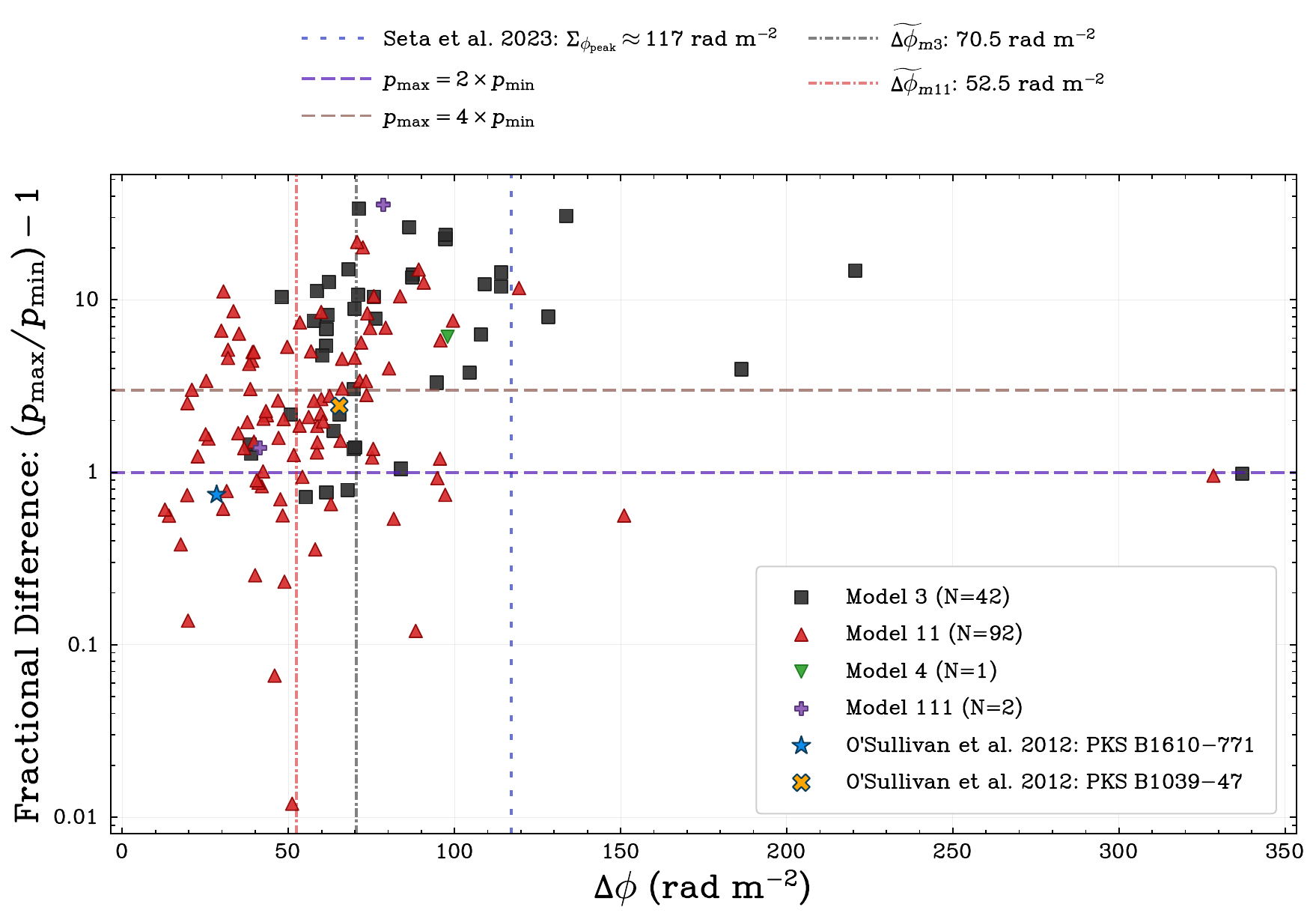}
    \caption{Difference in the fractional polarisation of each component for multi-component models against their difference in Faraday depth. For two-component models (m3, m4, m11), $\Delta\phi= \lvert\phi_1-\phi_2\rvert$ and the fractional polarisation difference is calculated as $(p_\mathrm{max}/p_\mathrm{min})-1$. For the three-component model 111, $\Delta\phi$ represents the mean pairwise difference among the three components: $\Delta\phi_{\mathrm{m}111} = \left\langle \lvert \phi_i - \phi_j \rvert \right\rangle_{\substack{i<j}}$, and the fractional polarisation difference is the mean of the three pairwise ratios minus 1: $\left\langle (p_i/p_j) - 1 \right\rangle_{i<j}$. Median lines of $\widetilde{\Delta\phi}_{\mathrm{m3}}$ and $\widetilde{\Delta\phi}_{\mathrm{m11}}$ are shown as dot-dashed black and red lines respectively. The expected global turbulence for the SMC from \citealt{SetaEA2023} at approximately $\Sigma_{\phi_\mathrm{peak}} \approx 117 ~\mathrm{rad~m}^{-2}$ is displayed as the blue dashed line. \citealt{2012.O’Sullivan} identified the two-component AGN PKS B1610-771 (blue star), along with a three-component AGN PKS B1039-47 (orange cross). Reference lines of two and four times the difference in fractional polarisation between components are shown as the purple and brown dashed lines respectively. Sources lower in fractional differences have components similar in polarised insensitivity.}
    \label{fig: delta_RM_vs_fracPol_ratio_model_3_11_4_111}
\end{figure}

The appearance of Faraday complexity in the sample is likely a result \revb{of} source structure, reaffirmed in Figure~\ref{fig: delta_RM_vs_fracPol_ratio_model_3_11_4_111}, which examines the fractional difference in fractional polarised intensity ($(p_\mathrm{max}/p_\mathrm{min})-1$) and difference in Faraday depth ($\Delta\phi= \lvert\phi_1-\phi_2\rvert$) between the components of models m3, m4, m11 and m111. If the multi-component structure arose from SMC turbulence, we would expect component separations to follow predictions based on the SMC's magneto-ionic properties. \cite{SetaEA2023} extrapolated the RM standard deviation at $\Sigma_{\phi_\mathrm{peak}} \approx 117 ~\mathrm{rad~m}^{-2}$ with RM structure functions using the data of \cite{2022MNRAS.510..260LivingstonSMC}, at the break scale of 160 pc. This prediction is shown as the blue dashed line in Figure~\ref{fig: delta_RM_vs_fracPol_ratio_model_3_11_4_111} and represents the expected $\Delta\phi$ for a second component if induced by global SMC turbulence. However, our observed component separations do not follow this prediction, but rather follow an uncorrelated scatter, suggesting that the multi-component structure is independent of the SMC magneto-ionic medium. Within this field, we capture more RM measurements across the entire Magellanic ecosystem than ever before. However, there exists the possibility that separating sources on and off the SMC is not feasible if the Magellanic ecosystem contains ionised Faraday-rotating gas beyond the ionised halo of the SMC, extending towards the Stream, Bridge, and Bar of the ecosystem. If this gas were Faraday-rotating and detectable at these frequencies, the Faraday dispersion observed in Model 2 from $QU$-fitting could arise from an extended envelope of magnetized gas beyond the local boundaries of the SMC. Additionally, the extent of interaction with the Milky Way's halo remains unclear \citep{2016.DOnghia, 2022.Choi, 2024ApJ...976L..28Mishra, 2025ApJ...983..151Carr}. Hence, with our SMC data alone, we cannot decisively distinguish the astrophysical origin of Faraday complexity.

More broadly, the physical origin of multiple Faraday components is not always clear. They may arise from intrinsic source structure, magneto-ionic material local to the source environment, or intervening foreground media \citep{1984ARA&A..22..319B, 1984ApJS...54..291P, 2017.O'Sullivan}. In the case of Figure~\ref{fig: delta_RM_vs_fracPol_ratio_model_3_11_4_111}, one might ideally expect distinct clustering or correlations in this parameter space that could help disentangle their origin. For instance, the Laing-Garrington effect predicts systematic asymmetries in depolarisation between the approaching and receding lobes of double radio sources \citep{1988Natur.331..147G, 1988Natur.331..149Laing}, while sources of similar morphology, such as radio galaxies or AGN with unresolved bipolar jets aligned along the LOS might be expected to exhibit comparable properties in $\Delta\phi$ and fractional polarisation difference, modulated by inclination angle \citep{2017.O'Sullivan}. In a sufficiently large and representative population sample drawn from full-sky observations, sources of distinct morphological and astrophysical types may therefore occupy well-defined regions within this parameter space. Additionally, the observed Faraday depth of a source is subject to cosmological $(1+z)^{-2}$ dilution, meaning that intrinsic Faraday rotation at the source redshift is reduced in the observer frame, introducing further scatter for sources at different redshifts \citep{2008.Kronberg, 2013.Durrer}. One might also expect the large-scale group or cluster environment to contribute additional Faraday complexity for sources embedded in dense magneto-ionic media, though we assume this is not a dominant effect for our sample. Beyond this, small-scale Faraday rotation fluctuations from the Milky Way on angular scales comparable to the data must also be considered. With the deeper fields and increased broadband frequency coverage that are becoming prevalent in the radio community, particularly with the advent of the SKA, filling in this parameter space becomes a realistic possibility. 


\subsubsection{Identification of Faraday Complexity}

\begin{figure}
    \centering
    \includegraphics[width = \columnwidth]{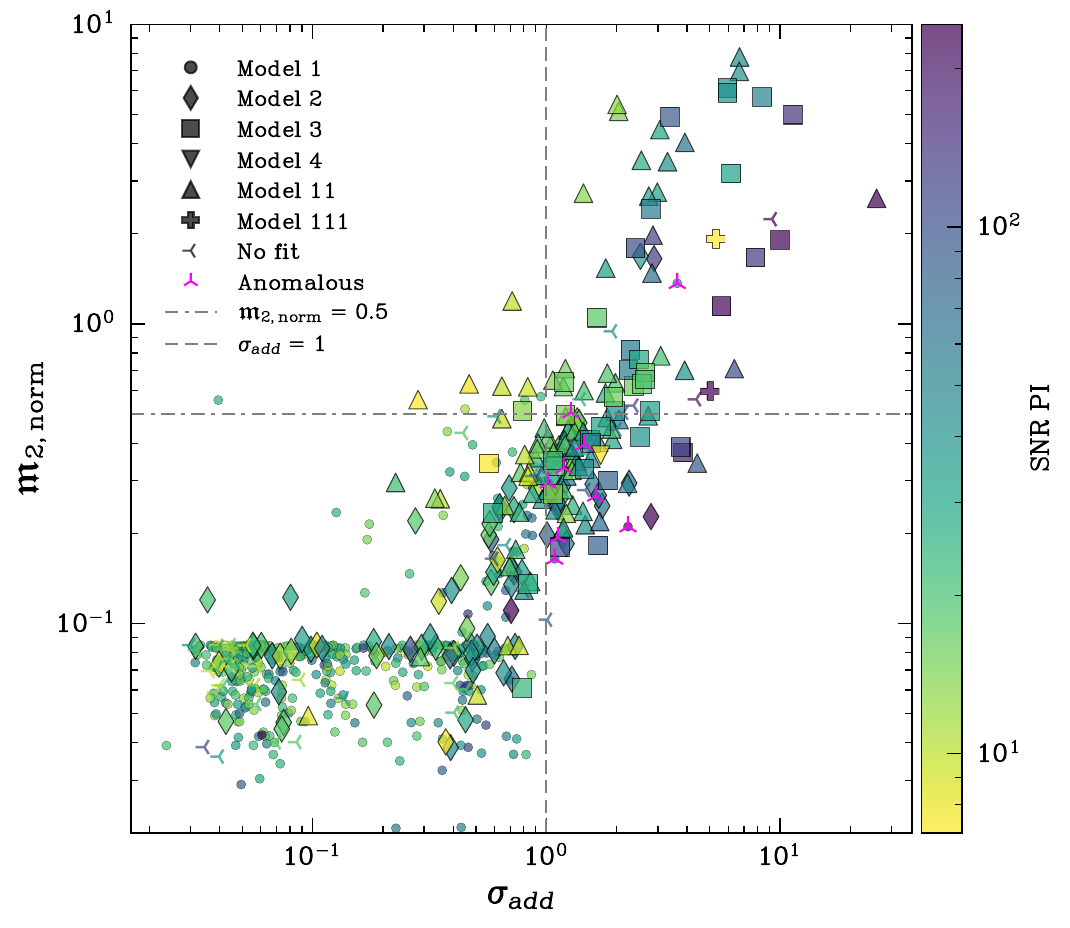}
    \caption{Normalised second moment ($\mathfrak{m}_{2,\mathrm{norm}}$) versus $\sigma_{\mathrm{add}}$ for all polarised sources, coloured by \revb{their PI SNR}. Sources are distinguished by their best-fit $QU$-fitting model: circles for Model 1, diamonds for Model 2, squares for Model 3, triangles for Model 11, and crosses for Model 111. Magenta coloured tri markers indicate sources designated as model 1 with a $\sigma_{\mathrm{add}}>1$ these sources removed from the catalogue are discussed in Section \ref{sec: Farady complexity discussion}. Horizontal and vertical lines indicate the thresholds for Faraday complexity classification, set at $\mathfrak{m}_{2,\mathrm{norm}} = 0.5$ and $\sigma_{\mathrm{add}} = 1$. 85\% of sources are within the lower left quadrant.}
    \label{fig: m2 vs sigma add}
\end{figure}

The motivation for Faraday complexity classification arises from the need to investigate the Faraday-rotating plasmas and in our quantification of the $\phi$. In Figure~\ref{fig: m2 vs sigma add}, we present the two Faraday complexity metrics derived from RM synthesis, $\mathfrak{m}_{2,\mathrm{norm}}$ and $\sigma_{\mathrm{add}}$, for all polarised sources; these sources are separated according to their best-fit model obtained from $QU$-fitting. Using complexity thresholds suitable for POSSUM data \cite[see Section~4.3.4 in][]{2024.Vanderwoude}, namely $\mathfrak{m}_{2,\mathrm{norm}} = 0.5$ and $\sigma_{\mathrm{add}} = 1$, the lower-left quadrant of the figure corresponds to sources classified as Faraday simple by both metrics, while the upper-right quadrant corresponds to sources deemed Faraday complex by both metrics.

When holding $QU$-fitting as the ground truth, several sources are misclassified by both metrics; however, the misclassification rate appears to be higher for $\mathfrak{m}_{2,\mathrm{norm}}$ than for $\sigma_{\mathrm{add}}$. The \revb{PI SNR} of the source is also crucial for correct classification amongst the RM synthesis tools. For sources with SNR PI $\ge 94$, $\sigma_{\mathrm{add}}$ correctly classify 88\%, compared to 38\% for $\mathfrak{m}_{2,\mathrm{norm}}$. To reach 90\% accuracy, $\sigma_{\mathrm{add}}$ requires an SNR PI $\ge 118$, while $\mathfrak{m}_{2,\mathrm{norm}}$ requires SNR PI $\ge 772$. However, at low \revb{PI SNR}, $QU$-fitting is also biased toward simpler models, so the trend of increasing agreement at higher \revb{PI SNR} cannot be attributed solely to improved RM synthesis metric performance.

While $\sigma_\mathrm{add}$ \revb{was} an effective diagnostic of strongly Faraday-complex behaviour with moderate \revb{PI SNR}, it \revb{was} less sensitive to mild complexity. Consequently, sources exhibiting high $\sigma_\mathrm{add}$ that \revb{were} classified as Model 1 (Faraday simple) by $QU$-fitting, indicating Faraday complexity that \revb{could not} be captured by our multi-component models. We \revb{identified} nine such sources in the field and \revb{excluded} them from further analysis, as no suitable model adequately \revb{described} their polarisation behaviour. These sources are marked as anomalous in Figures~\ref{fig: RMs from QU fitting in the sky} and \ref{fig: m2 vs sigma add}. One example exhibiting repolarisation is presented in Appendix~\ref{sec: Unique sources}. The spectral characteristics of these anomalous sources likely \revb{arose} from intrinsic structure within the emitting volume \citep{2015.Anderson}, which our externally screened models \revb{could not} reproduce.

The \rev{116} sources that were detected in RM synthesis but failed the $QU$-fitting reliability criteria were too faint, with median \revb{SNR Stokes $I$} and \revb{PI SNR} approximately half those of successfully fitted sources. Despite being intrinsically polarised (median fractional polarisation of 6.3\%), their low flux densities precluded reliable model fitting. The spatial distribution is random across the field (see Figure~\ref{fig: RMs from QU fitting in the sky}).

\bsp	
\label{lastpage}

\end{document}